%% file: main_plb_v2.tex
\documentclass[%
 reprint,
superscriptaddress,
nofootinbib,
 amsmath,amssymb,
 aps,
 prd,
]{revtex4-2}

\usepackage{graphicx}
\usepackage{tikz-feynhand}
\usepackage{hyperref}
\usepackage{physics}
\usepackage{enumitem}
\usepackage{comment}

\newcommand{\lam}{\lambda}
\newcommand{\eps}{\epsilon}

\newcommand{\rev}[1]{{\color{black}#1}}

\begin{document}


\title{Higgs production in association with a Z boson\\ at TeV-scale lepton colliders}


\author{Hiroyuki Furusato}
\affiliation{Graduate School of Science and Engineering, Iwate University, Morioka, Iwate 020-8550, Japan}
\author{Satsuki Hosoya}
\affiliation{Graduate School of Arts and Sciences, Iwate University, Morioka, Iwate 020-8550, Japan}
\author{Kentarou Mawatari}
\affiliation{Graduate School of Science and Engineering, Iwate University, Morioka, Iwate 020-8550, Japan}
\affiliation{Graduate School of Arts and Sciences, Iwate University, Morioka, Iwate 020-8550, Japan}
\affiliation{Faculty of Education, Iwate University, Morioka, Iwate 020-8550, Japan}
\author{Shouta Suzuki}
\affiliation{Graduate School of Arts and Sciences, Iwate University, Morioka, Iwate 020-8550, Japan}



\begin{abstract}
We study kinematical properties of the process $l^-l^+ \to \nu\bar{\nu}Zh$ at future TeV-scale lepton colliders.
We classify the Feynman amplitudes into three main topological categories: vector-boson scattering, $l^-W^+$ scattering, and $W^-l^+$ scattering contributions,
and analyze the interference patterns among these amplitudes in both the recently proposed Feynman-diagram (FD) gauge and the conventional unitary gauge.
We demonstrate that the rapidity distributions of the $Z$ boson closely resemble those in $l^-l^+ \to \nu\bar{\nu}Z$, while the Higgs-boson rapidity distribution is significantly different: Higgs bosons are produced nearly isotropically, whereas longitudinally polarized $Z$ bosons are enhanced in the forward and backward regions.
We find that this difference is traced to the topology-dependent structure of the FD-gauge amplitudes.
We also demonstrate that suitable kinematical cuts can isolate specific subamplitudes, enabling partial FD-gauge amplitudes to effectively describe physical distributions. 
\end{abstract}


\maketitle

\section{Introduction}\label{sec:intro}

Associated production of a weak vector boson with a Higgs boson ($Vh$ production) via vector boson scattering (VBS) is known as an important process 
\rev{
in the Higgs precision program
}
for the current and future high-energy collider experiments 
since the cross section becomes larger as the collision energy increases.
In addition, the process includes diagrams where the Higgs boson couples to either a W or Z boson, 
and hence it is sensitive to physics related to electroweak symmetry breaking.
The opposite-sign coupling hypothesis between the $hWW$ and $hZZ$ couplings is already excluded with significance beyond 5$\sigma$ by the VBS $Wh$ production at the LHC~\cite{ATLAS:2024vxc,CMS:2024srp}.
The relative sign of the couplings in this process for the LHC~\cite{deLima:2021llm,deLima:2024uwc} as well as  
for future lepton colliders~\cite{Stolarski:2020qim,deLima:2024ybb} has also been studied. 

However, despite being important as mentioned above, 
the kinematical distributions of such VBS $Vh$ production processes, even in the standard model (SM), have not been extensively studied. 
In this paper, therefore, we consider the process
\begin{align}
  l^-l^+\to \nu_l\bar{\nu}_lZh 
\label{ll_vvzh}    
\end{align}
in the SM in order to study the kinematical properties in detail
for future TeV-scale lepton colliders~\cite{LinearColliderVision:2025hlt},
such as the higher-energy stage of the International Linear Collider (ILC)~\cite{ILCInternationalDevelopmentTeam:2022izu}, the Compact Linear Collider (CLIC)~\cite{Roloff:2018dqu}, and the international muon collider~\cite{InternationalMuonCollider:2024jyv}. 
As shown below, the kinematical distributions for the longitudinally-polarized Z boson and the Higgs boson in the process~\eqref{ll_vvzh} are rather different,
contrary to our naive expectations. 
This paper aims to explore the reasons behind this difference. 

It should be noted here that 
one reason the process~\eqref{ll_vvzh} has not been studied much so far is 
 a problem of numerical simulation in the energy region above a few TeV
because of subtle gauge cancellation among the amplitudes in the conventional unitary (U) gauge.%
\footnote{
The default {\tt MadGraph5\_aMC@NLO} ({\tt MG5aMC} henceforth) v3.7.0~\cite{Alwall:2014hca} can generate only a few hundred events 
for $e^-e^+\to \nu_e\bar{\nu}_eZh$ in the energy range above 3~TeV despite the request for 10,000 event generation.
Note that the default gauge choice for tree-level processes in {\tt MG5aMC} is the U gauge for massive gauge bosons.
Such problems of VBS-like processes can be solved by a different phase-space integration strategy~\cite{Mattelaer:2021xdr}.
}
As a solution to avoid such a numerical problem, the Feynman-diagram (FD) gauge has recently been proposed~\cite{Hagiwara:2020tbx,Chen:2022gxv,Chen:2022xlg}, and 
was implemented in {\tt MG5aMC}~\cite{Hagiwara:2024xdh}, so that an arbitrary tree-level process for an arbitrary gauge model can be studied in the FD gauge. 
The studies for similar processes such as $l^-l^+\to \nu\bar{\nu}VV$ for $V=W^\pm$ and/or $Z$~\cite{Chen:2022gxv} and $l^-l^+\to \nu\bar{\nu}t\bar th$~\cite{Hagiwara:2024xdh} have been reported at the level of total cross sections
to demonstrate that, unlike the U gauge, FD-gauge amplitudes are free from subtle gauge cancellation among interfering amplitudes at high energies.
Moreover, Refs.~\cite{Chen:2022gxv,Chen:2022xlg,Furusato:2024ghr,Hagiwara:2026lul} studied the interference patterns among the FD-gauge amplitudes in kinematical distributions for electroweak processes,
and showed that the physical distributions can be interpreted by the contributions from the individual amplitudes. 
Similar studies can also be found in Refs.~\cite{Jeong:2025hwj,Li:2025ikn}. 

In this paper, 
we study kinematical properties of the process~\eqref{ll_vvzh} in detail, 
and show that the difference in distributions between the Z boson and the Higgs boson is traced to the topology-dependent structure of the FD-gauge amplitudes.
In addition, we also demonstrate that suitable kinematical cuts can isolate specific subamplitudes, enabling partial FD-gauge amplitudes to effectively describe physical distributions. 

In the process of understanding the interference patterns among the amplitudes for the process~\eqref{ll_vvzh}, 
we find the similarity in the distributions of the Z boson in the process~\cite{Hagiwara:1990gk}
\begin{align}
  l^-l^+\to \nu_l\bar{\nu}_lZ\ . 
\label{ll_vvz}  
\end{align}
Since it is easier to understand the properties of the amplitudes in the process~\eqref{ll_vvz}
as the number of the Feynman diagrams is less than in the process~\eqref{ll_vvzh}, 
we first analyze the process~\eqref{ll_vvz} in detail, and study the process~\eqref{ll_vvzh} later in this article. 

The paper is organized as follows. 
In Sec.~\ref{sec:fd} we briefly introduce the FD gauge.
In Sec.~\ref{sec:xsec} we present total cross sections for the processes~\eqref{ll_vvzh} and \eqref{ll_vvz}.
In Sec.~\ref{sec:vvz}, 
we study the $l^-l^+\to \nu\bar{\nu}Z$ process rather in detail. 
We show the total and differential cross sections to discuss behavior of the FD-gauge amplitudes through the comparison with those in the U gauge. 
Then, we investigate the interference patterns among the amplitudes.
In Sec.~\ref{sec:vvzh} we repeat a similar study for $l^-l^+\to \nu\bar{\nu}Zh$ 
with emphasis on the difference of the kinematical distributions between the Z boson and the Higgs boson. 
Section~\ref{sec:summary} summarizes our findings.
\rev{
Although we focus on rapidity distributions in the main text, 
we also present transverse-momentum distributions in Appendix as a supplement. 
}

\section{Feynman-diagram gauge}\label{sec:fd}

In this section, we briefly introduce the FD gauge. See Refs.~\cite{Chen:2022gxv,Chen:2022xlg} for more details.

In the FD gauge, the propagator of the gauge bosons is obtained from the light-cone gauge~\cite{Chen:2022xlg}, 
where the gauge vector is chosen along the opposite direction of the gauge-boson three momentum
\begin{align}
n^\mu = ( {\rm sgn}(q^0), -\vec{q}/|\vec{q}| )\ .
\label{eq:FD}
\end{align}

The equation of motion for massive weak bosons mixes the four-component weak boson vector $V^\mu$ and the associated Goldstone boson $\pi_V$, 
making them five component fields and their $5\times5$ matrix propagators~\cite{Chen:2022gxv,Chen:2022xlg},
\begin{align}
G_{MN}(q) = \frac{i}{q^2-m^2+i\epsilon} 
  \begin{pmatrix}
          -g_{\mu\nu} 
          + \dfrac{q_{\mu}n_{\nu}+n_{\mu}q_{\nu}  }{n\cdot q } & i \dfrac{m\, n_\mu}{n\cdot q} \\
          -i \dfrac{m\, n_\nu}{n\cdot q} & 1 \\
  \end{pmatrix} 
\label{fdpropagator2}
\end{align}
with $M,N=0$ to 4, where 0 to 3 are the Lorentz indices $\mu,\nu$.

The polarization vector (or wave function) for massive gauge bosons with the helicity $\lam\,(=\pm1,0)$ in the FD gauge is given as a five-component vector by~\cite{Chen:2022gxv}
\begin{align}
 \eps^M(q,\pm)&=(\eps^\mu(q,\pm),\,0)\ , 
 \label{fdpolT}\\
 \eps^M(q,0)&=(\tilde\eps^\mu(q,0),\,i)\ ,
 \label{fdpolL}
\end{align}
with the reduced polarization vector~\cite{Hagiwara:2020tbx,Chen:2022gxv,Chen:2022xlg}
\begin{align}
 \tilde\eps^\mu(q,0)&=\eps^\mu(q,0)-\frac{q^\mu}{Q}
  =-{\rm sgn}(q^2)\frac{Q\,n^\mu}{n\cdot q} \ ,
 \label{e0tilde}
\end{align}
where $\eps^\mu(q,\lam)$ is the ordinary polarization vector and $Q=\sqrt{|q^2|}$.

\section{Total cross sections}\label{sec:xsec}

We start by showing the energy dependence of total cross sections for
 $\nu\bar{\nu}Z$ and $\nu\bar{\nu}Zh$ productions at lepton colliders.
All numerical results in this paper are at the tree level in the SM, and 
are obtained by employing {\tt MG5aMC}~\cite{Alwall:2014hca} 
with the FD-gauge implementation~\cite{Hagiwara:2024xdh}.%
\footnote{
\rev{
A command ``{\tt set gauge FD}" to use the FD gauge is available in the {\tt MG5aMC} version of 3.6.0 or higher.
}
}
\rev{
We calculate all physical observables, i.e. total and differential cross sections, independently in both the FD gauge and the U gauge, and confirm that those agree within numerical precision to validate the program.
}

Contributions from the $l^-l^+\to Z$ annihilation ($l=e$ or $\mu$) are significant only when $\sqrt{s}<1$~TeV 
and negligible above multi-TeV energy regions for both processes~\cite{Furusato:2026vhc}.
Therefore, to make physics discussion simpler, we consider $e^-\mu^+$ collisions in the following as
\begin{align}
  e^-\mu^+\to \nu_{e}\bar{\nu}_{\mu}Z\ , 
\label{vvz}
\end{align}
and
\begin{align}
  e^-\mu^+\to \nu_e\bar{\nu}_\mu Zh\ ,
\label{vvzh}
\end{align}
where only $t$-channel W-boson exchange diagrams contribute;
the explicit Feynman diagrams are shown in Sec.~\ref{sec:vvz} and Sec.~\ref{sec:vvzh}, respectively.
We note that only the left-handed electron and the right-handed $\mu^+$ collision, $e_L^-\mu_R^+$, gives non-zero cross section for the processes \eqref{vvz} and \eqref{vvzh}.

\begin{figure}
\centering
\includegraphics[width=1\columnwidth]{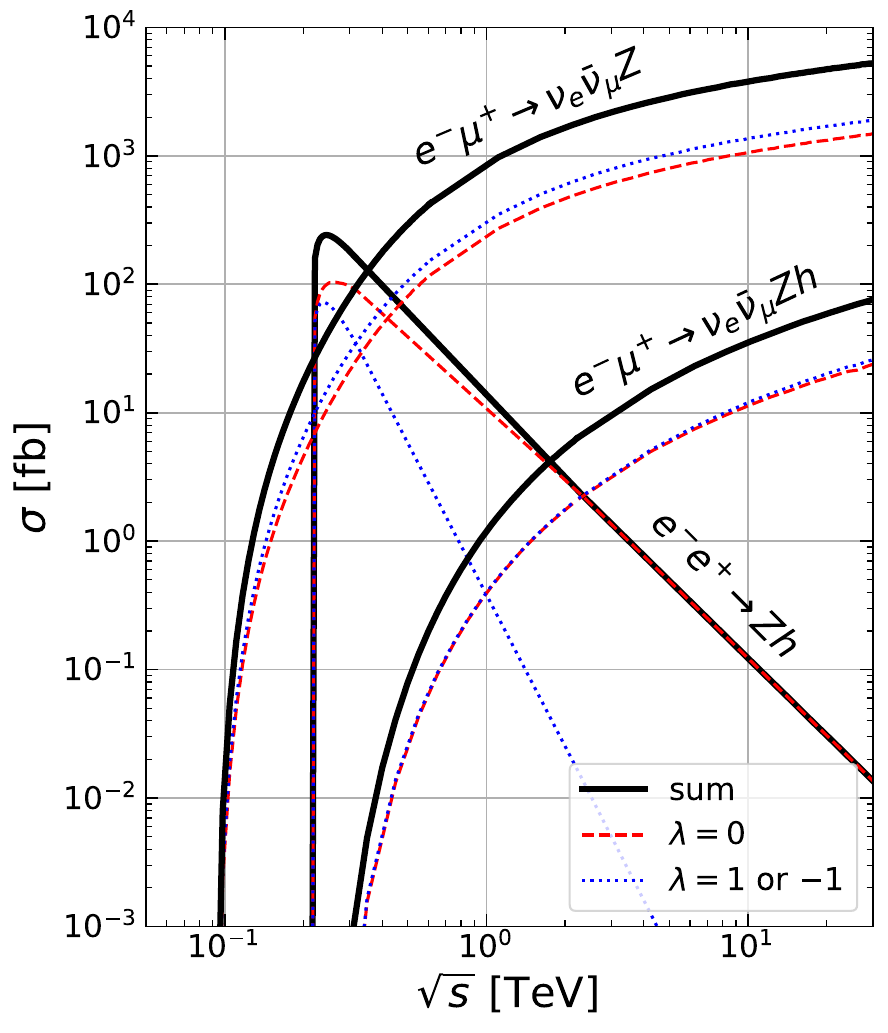}
\caption{
Total cross sections of $e^{-}\mu^{+}\to\nu_e\bar\nu_\mu Z$ and $e^{-}\mu^{+}\to \nu_e\bar{\nu}_\mu Zh$ as a function of the collision energy.
The helicity-dependent cross sections for $\lambda=0$ and $\pm1$ are also shown by red-dashed and blue-dotted lines, respectively,  
where $\lambda$ is the helicity of the Z boson in the $l^-l^+$ collision c.m. frame.
Total cross sections of $e^{-}e^{+}\to Zh$ are also shown as a reference.
}
\label{fig:xsec}
\end{figure}

In Fig.~\ref{fig:xsec}, as a reference, we also show the total cross section for $e^{-}e^{+}\to Zh$, which
has a peak at around $\sqrt{s}=250$~GeV and falls as $1/s$ at high energies.
In contrast, 
the cross sections for both $\nu\bar{\nu}Z$ and $\nu\bar{\nu}Zh$ productions
grow as the collision energy increases, and
become larger than the $Zh$ cross section when $\sqrt{s}\gtrsim400$~GeV and $\sqrt{s}\gtrsim2$~TeV, respectively.
Both $\nu\bar{\nu}Z$ and $\nu\bar{\nu}Zh$ processes are via the $t$-channel W-boson exchanges,
and hence they share the same logarithmic structure,
leading to very similar energy dependence of the cross sections, differing mainly by an overall normalization.

Figure~\ref{fig:xsec} also shows the helicity-dependent cross sections, 
where $\lambda$ is the helicity of the Z boson in the $l^-l^+$ collision c.m. frame.
We find that for the $Zh$ production the longitudinally-polarized ($\lambda=0$) Z boson, denoted by a red-dashed line, is dominantly produced in high energies.
For the $\nu\bar\nu Z$ and $\nu\bar\nu Zh$ productions, on the other hand, 
the production rates for $\lambda=0$ (red-dashed) and $\lambda=\pm1$ (blue-dotted) are very similar in all energy regions. 
Note that the total cross sections for $\lambda=+1$ and for $\lambda=-1$ Z bosons are identical. 

In the following, we focus on the production of the longitudinally-polarized ($\lambda=0$) Z boson.
The same basic arguments apply to the transverse case ($\lambda=\pm1$), which we will report separately. 
\rev{
We note that realistic analyses for experiments require decays of Z and Higgs bosons,
where all the polarized states of the Z bosons should be taken into account~\cite{Ballestrero:2019qoy}, 
as well as detector effects.
}

\section{$l^-l^+\to \nu\bar{\nu}Z$}\label{sec:vvz}

\begin{figure}[b]
\centering
\begin{tikzpicture}[scale=0.6]
\begin{feynhand}
    \vertex [particle] (i1) at (-1,-2) {$e^{-}$};
    \vertex [particle] (i2) at (1,-2) {$\mu^{+}$};
    \vertex [particle] (f1) at (-1,2) {$\nu_{e}$};
    \vertex [particle] (f2) at (1,2) {$\bar{\nu}_{\mu}$};
    \vertex (a) at (0,0);
    \vertex [particle] (z) at (0,2) {$Z$};
    \vertex (w1) at (-1,0);
    \vertex (w2) at (1,0);
    \propag [plain] (i1) to (w1);
    \propag [plain] (w1) to (f1);
    \propag [plain] (i2) to (w2);
    \propag [plain] (w2) to (f2);
    \propag [boson] (w1) to [edge label'=$W$] (a);
    \propag [boson] (w2) to [edge label=$W$](a); 
    \propag [boson] (a) to (z);
\end{feynhand}
\end{tikzpicture}
\begin{tikzpicture}[scale=0.6]
\begin{feynhand}
    \vertex [particle] (i1) at (-0.5,-2) {$e^{-}$};
    \vertex [particle] (i2) at (0.5,-2) {$\mu^{+}$};
    \vertex [particle] (f1) at (-0.5,2) {$\nu_{e}$};
    \vertex [particle] (f2) at (0.5,2) {$\bar{\nu}_{\mu}$};
    \vertex [particle] (a) at (-0.5,-1);
    \vertex [particle] (z) at (-1.2,0.25) {$Z$};
    \vertex (w1) at (-0.5,0);
    \vertex (w2) at (0.5,0);
    \propag [plain] (i1) to (w1);
    \propag [plain] (w1) to (f1);
    \propag [plain] (i2) to (w2);
    \propag [plain] (w2) to (f2);
    \propag [boson] (w1) to [edge label'=$W$] (w2);
    \propag [boson] (a) to (z);
\end{feynhand}
\end{tikzpicture}
\begin{tikzpicture}[scale=0.6]
\begin{feynhand}
    \vertex [particle] (i1) at (-0.5,-2) {$e^{-}$};
    \vertex [particle] (i2) at (0.5,-2) {$\mu^{+}$};
    \vertex [particle] (f1) at (-0.5,2) {$\nu_{e}$};
    \vertex [particle] (f2) at (0.5,2) {$\bar{\nu}_{\mu}$};
    \vertex [particle] (a) at (-0.5,0.75);
    \vertex [particle] (z) at (-1.2,2) {$Z$};
    \vertex (w1) at (-0.5,0);
    \vertex (w2) at (0.5,0);
    \propag [plain] (i1) to (w1);
    \propag [plain] (w1) to (f1);
    \propag [plain] (i2) to (w2);
    \propag [plain] (w2) to (f2);
    \propag [boson] (w1) to [edge label'=$W$] (w2);
    \propag [boson] (a) to (z);
\end{feynhand}
\end{tikzpicture}
\begin{tikzpicture}[scale=0.6]
\begin{feynhand}
    \vertex [particle] (i1) at (-0.5,-2) {$e^{-}$};
    \vertex [particle] (i2) at (0.5,-2) {$\mu^{+}$};
    \vertex [particle] (f1) at (-0.5,2) {$\nu_{e}$};
    \vertex [particle] (f2) at (0.5,2) {$\bar{\nu}_{\mu}$};
    \vertex [particle] (a) at (0.5,-1);
    \vertex [particle] (z) at (1.2,0.25) {$Z$};
    \vertex (w1) at (-0.5,0);
    \vertex (w2) at (0.5,0);
    \propag [plain] (i1) to (w1);
    \propag [plain] (w1) to (f1);
    \propag [plain] (i2) to (w2);
    \propag [plain] (w2) to (f2);
    \propag [boson] (w1) to [edge label'=$W$] (w2);
    \propag [boson] (a) to (z);
\end{feynhand}
\end{tikzpicture}
\begin{tikzpicture}[scale=0.6]
\begin{feynhand}
    \vertex [particle] (i1) at (-0.5,-2) {$e^{-}$};
    \vertex [particle] (i2) at (0.5,-2) {$\mu^{+}$};
    \vertex [particle] (f1) at (-0.5,2) {$\nu_{e}$};
    \vertex [particle] (f2) at (0.5,2) {$\bar{\nu}_{\mu}$};
    \vertex [particle] (a) at (0.5,0.75);
    \vertex [particle] (z) at (1.2,2) {$Z$};
    \vertex (w1) at (-0.5,0);
    \vertex (w2) at (0.5,0);
    \propag [plain] (i1) to (w1);
    \propag [plain] (w1) to (f1);
    \propag [plain] (i2) to (w2);
    \propag [plain] (w2) to (f2);
    \propag [boson] (w1) to [edge label'=$W$] (w2);
    \propag [boson] (a) to (z);
\end{feynhand}
\end{tikzpicture}
\hspace*{0mm}(a)\hspace*{14mm}(b-i)\hspace*{10mm}(b-f)\hspace*{7mm}(c-i)\hspace*{10mm}(c-f)
\caption{Feynman diagrams for $e^-\mu^+\to \nu_{e}\bar{\nu}_{\mu}Z$.}
\label{fig:vvz}
\end{figure}
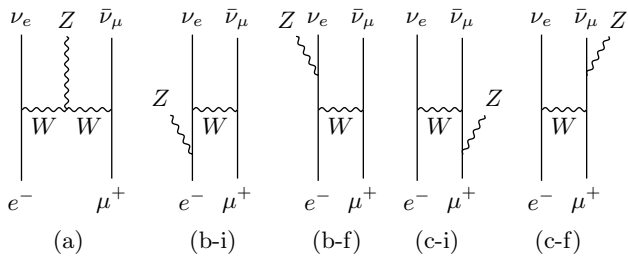

As mentioned in Sec.~\ref{sec:intro},  
we first study the $l^-l^+\to \nu\bar{\nu}Z$ process~\eqref{vvz} rather in detail in this section
in order to understand the interference patterns among the amplitudes for the $l^-l^+\to \nu\bar{\nu}Zh$ process~\eqref{vvzh},
which is the main target in this article.
We classify the amplitudes of the process~\eqref{vvz} into a few groups,
and show the total and differential cross sections to discuss the gauge dependence of the contributions from each group through the comparison between the FD and U gauges. 

There are five Feynman diagrams for the process~\eqref{vvz}, 
depending on where a Z boson is emitted from the $e^-\mu^+\to\nu_e\bar\nu_\mu$ process,
 as shown in Fig.~\ref{fig:vvz}.
We classify the Feynman amplitudes into three main groups based on the topology of each Feynman diagram~\cite{Hagiwara:2024xdh,Hagiwara:2026lul};
\begin{enumerate}
\setlength{\itemsep}{0cm} 
 \item[(a)] vector boson fusion (VBF),
 \item[(b)] $e^-W^+$ scattering,  
 \item[(c)] $W^-\mu^+$ scattering.
\end{enumerate}
The categories (b) and (c) each have two diagrams as
\begin{enumerate}[
  label={},
  leftmargin=!,
  labelwidth=2.2em,
  labelsep=0.5em,
  align=left
]
\setlength{\itemsep}{0cm} 
  \item[ (b-i)] $Z$ emission from the initial electron line ($e^-$),
  \item[ (b-f)] $Z$ emission from the final electron line ($\nu_e$),
  \item[ (c-i)] $Z$ emission from the initial muon line ($\mu^+$), 
  \item[ (c-f)] $Z$ emission from the final muon line ($\bar\nu_\mu$).
\end{enumerate}

Needless to say, physical cross sections, obtained by squaring the sum of all relevant amplitudes, are gauge invariant,
while contributions grouped according to the topology of each Feynman diagram are gauge dependent and unphysical.
Nevertheless, we study behavior of each group to demonstrate that 
physical distributions can be interpreted by the contributions from each subgroup in the FD gauge,
as shown in the previous studies~\cite{Hagiwara:2020tbx,Chen:2022gxv,Chen:2022xlg,Hagiwara:2024xdh,Furusato:2024ghr,Hagiwara:2026lul}.

\begin{figure}
  \includegraphics[width=1\columnwidth]{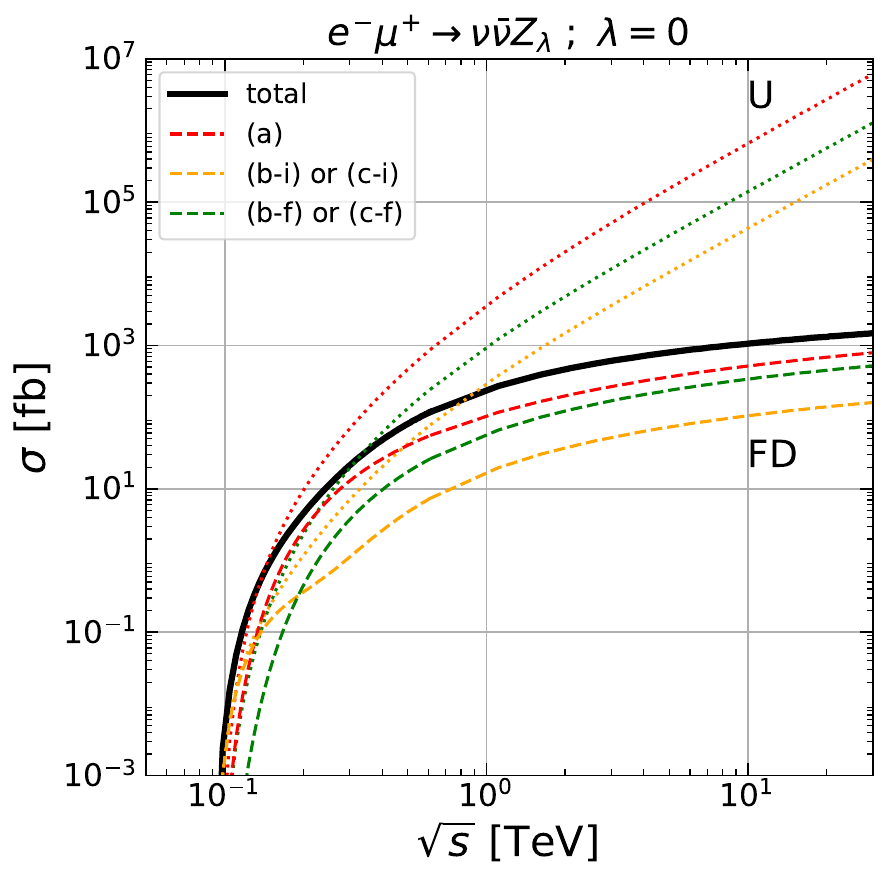}
  \caption{Total cross section of $e^{-}\mu^{+}\to \nu\bar{\nu}Z_{\lambda}$ for $\lambda=0$
as a function of the collision energy,
where $\lambda$ is the helicity of the Z boson in the $e^-\mu^+$ collision c.m. frame. 
The solid line denotes the total cross section, while dashed (dotted) lines show contributions from the absolute value squared of each amplitude, as defined in Fig.~\ref{fig:vvz}, in the FD (U) gauge.}
\label{fig:xsec_FDvsU_l0}
\end{figure}

Figure~\ref{fig:xsec_FDvsU_l0} shows the total cross section of $e^-\mu^+\to \nu_e\bar{\nu}_\mu Z$ for $\lambda=0$ as a function of the collision energy $\sqrt{s}$ from 100~GeV up to 30~TeV, where $\lambda$ is the helicity of the final-state $Z$ boson in the $e^-\mu^+$ collision c.m. frame. 
The black-solid line, which is identical to the red-dashed line for $e^-\mu^+\to \nu_e\bar{\nu}_\mu Z$ in Fig.~\ref{fig:xsec},
denotes the physical cross section, 
and does not depend on the gauge choice.
The contributions from each amplitude are shown by the dashed (dotted) lines in the FD (U) gauge.
The red, orange, and green lines show the contributions from (a) VBF, 
(b-i) or (c-i) $Z$ emission from the initial state, and (b-f) or (c-f) $Z$ emission from the final state, respectively. 
The lines of (b-i) and (c-i) as well as (b-f) and (c-f) completely overlap.

In the U gauge, shown by dotted lines, 
each contribution grows with energy as $(\sqrt{s})^2$ at high energies.
This is a well-known property of amplitudes for productions of the longitudinally-polarized gauge boson; see, e.g.~Ref.~\cite{Peskin:1995ev}.
We refer the readers to Ref.~\cite{Furusato:2024ghr} for analytic expressions of the amplitudes for the $l^-l^+\to W^-W^+$ process.
This growth of the amplitudes with energy requires significant cancellations among the individual contributions to obtain the physical cross section,
giving rise to a problem of numerical evaluation for the cross section especially at high energies,
as mentioned in Sec.~\ref{sec:intro}.

On the other hand, in the FD gauge, denoted by dashed lines, the contributions from each amplitude are along with the total cross section without any unphysical energy growth.
Unlike the U gauge, 
there is no artificial cancellation among the relevant amplitudes in the FD gauge. 

We note that the ratios of the $Z$-emission contribution from the final state (orange lines) to that from the initial state (green lines) 
are given by the ratio of the $Z$--neutrino coupling to the $Z$--charged-lepton coupling as
\begin{align}\label{eq:ratio}
  \frac{\sigma(\text{b-f})}{\sigma(\text{b-i})}
 =\frac{\sigma(\text{c-f})}{\sigma(\text{c-i})}
 \approx \left(\frac{g_{Z\nu\nu}}{g_{Zl_Ll_L}}\right)^2\sim 3.2
\end{align}
both in the FD and U gauges at high energies.

\begin{figure}
  \includegraphics[width=1\columnwidth]{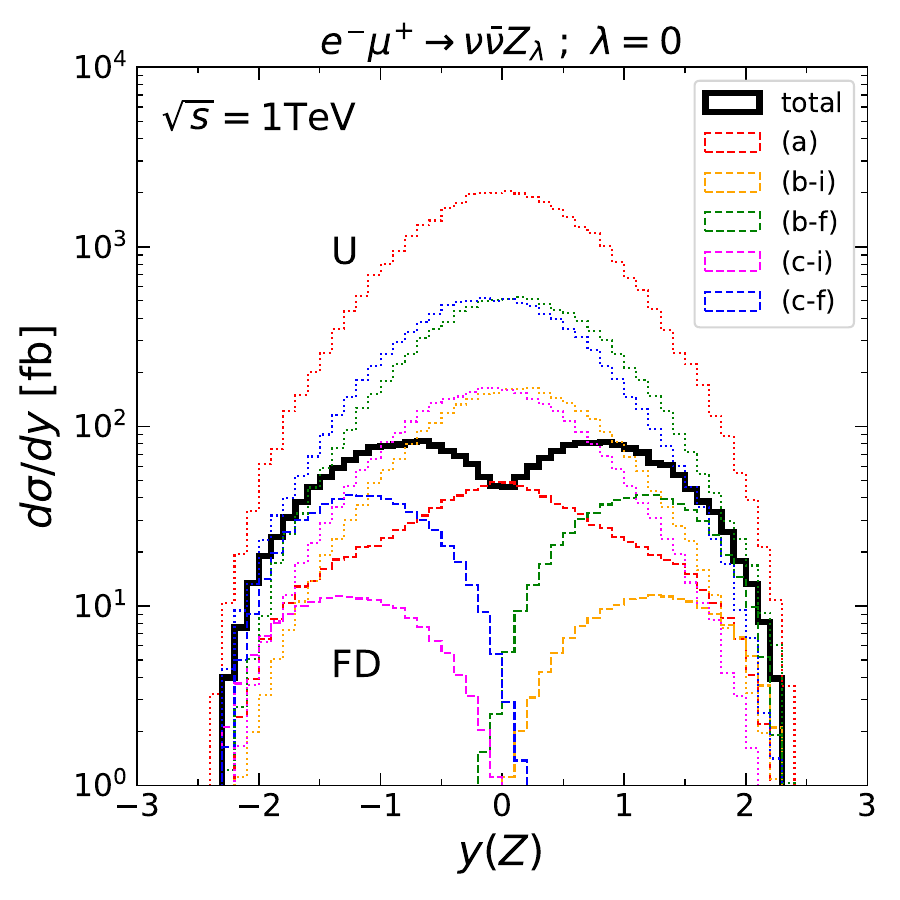}
  \caption{Rapidity distribution of the Z boson for $e^{-}\mu^{+}\to \nu\bar{\nu}Z_{\lambda}$ with $\lambda=0$ at $\sqrt{s}=1$~TeV,
where $\lambda$ is the helicity of the Z boson in the $e^-\mu^+$ collision c.m. frame. 
The solid line denotes the observable distribution, while dashed (dotted) lines show contributions from the absolute value squared of each amplitude, as defined in Fig.~\ref{fig:vvz}, in the FD (U) gauge.}
\label{fig:dxsec_FDvsU_l0}
\end{figure}

Let us move to kinematical distributions for $e^{-}\mu^{+}\to \nu\bar{\nu}Z$. 
\rev{
In the following, we focus on rapidity distributions in the main text, 
but see also Appendix for transverse-momentum distributions. 
}
Figure~\ref{fig:dxsec_FDvsU_l0} shows the rapidity distribution of the $Z$ boson for $\lambda = 0$ with the collision energy fixed at $1$~TeV. 
The line styles are the same as those in Fig.~\ref{fig:xsec_FDvsU_l0}, 
but the degeneracies between (b-i) and (c-i) as well as between (b-f) and (c-f) are resolved.

The black-solid line, i.e. the observable distribution, shows that the central Z-boson production ($y\sim0$) is suppressed and the distribution has  
peaks at around $|y|=0.8$.

In the U gauge (dotted lines), 
not only is the magnitude of the distributions much larger than the observable distribution, 
but the shapes are also entirely different, especially in the central region.
Therefore, the distributions from each amplitude 
do not provide any useful information. 
One can easily imagine that event generation with U-gauge amplitudes for such a process is very inefficient especially at high energies;
see the footnote in Sec.~\ref{sec:intro}.

In the FD gauge (dashed lines), in contrast, the properties of each Feynman diagram in Fig.~\ref{fig:vvz} apparently are reflected in the distributions. 
Note that we take the $e^-$ momentum in the initial state as the positive $z$-axis direction, while the $\mu^+$ is taken as the negative $z$-axis direction. 
The Z bosons produced via (a) VBF, denoted by the red-dashed line in Fig.~\ref{fig:dxsec_FDvsU_l0}, are 
distributed throughout the entire region and
produced more in the central region.
The distribution is symmetric under $y\leftrightarrow-y$.
On the other hand, the categories (b) $e^-W^+$ scattering and (c) $W^-\mu^+$ scattering have strong asymmetry in the rapidity distribution in the FD gauge.

\begin{figure*}
    \centering
    \includegraphics[width=0.495\textwidth]{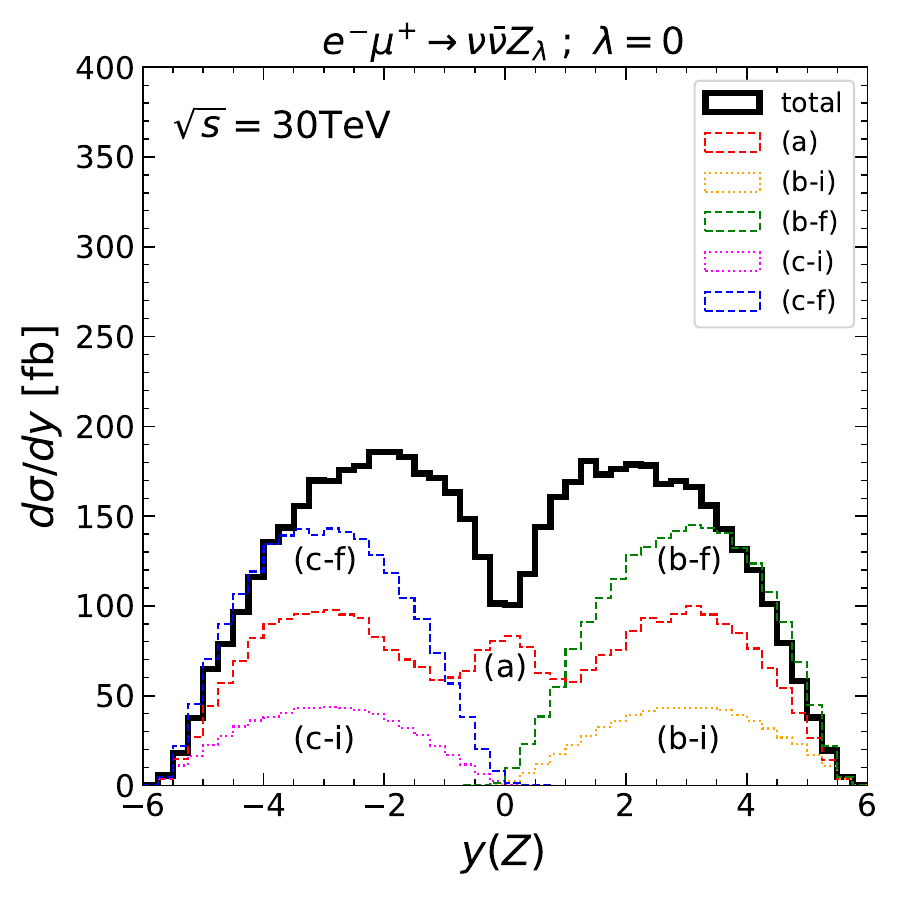}
    \includegraphics[width=0.495\textwidth]{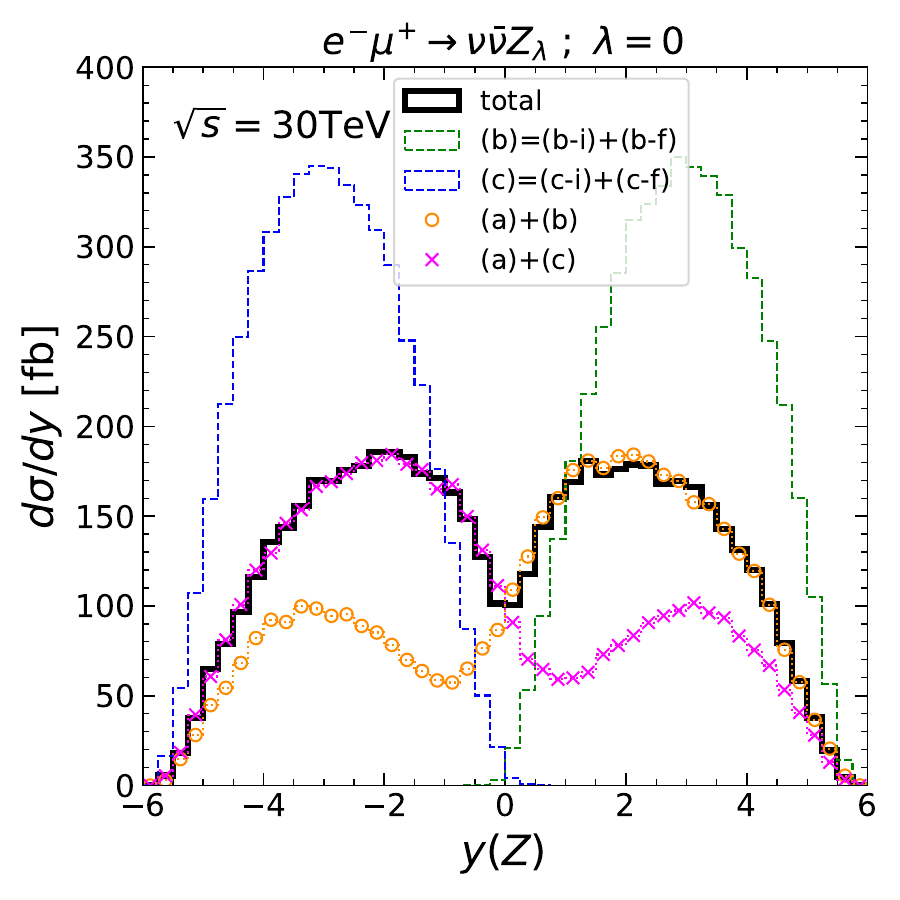}
    \caption{Rapidity distribution of the $Z$ boson for $e^{-}\mu^{+}\to \nu\bar{\nu}Z_{\lambda}$ with $\lambda=0$ at $\sqrt{s}=30\ \text{TeV}$. 
In both panels 
the solid lines denote the observable distribution. 
In the left panel the five individual contributions in the FD gauge are shown,
while in the right panel contributions from the absolute value squared of the sum of the two or three amplitudes are shown. 
	 }
    \label{fig:emuvvzh_yz_diagrams_if}
\end{figure*}

As expected from the $t$-channel topology of the Feynman diagrams in Fig.~\ref{fig:vvz}, 
the $\nu_e$ is scattered into the forward ($y>0$) region, 
while the $\bar\nu_\mu$ is scattered into the backward ($y<0$) region.
For (b) $e^-W^+$ scattering, the Z boson is emitted along the electron line,
and hence it tends to be scattered into the forward ($y>0$) region.
The green and orange dashed lines in Fig.~\ref{fig:dxsec_FDvsU_l0} are consistent with this expectation. 
Similarly, for (c) $W^-\mu^+$ scattering, 
the Z boson is emitted along the muon line,
and hence it tends to be scattered into the backward ($y<0$) region.
The blue and magenta dashed lines are consistent with this expectation.
We note that the two contributions from (b-i) and (c-i) are symmetric with each other under $y\to-y$. The same symmetry holds for (b-f) and (c-f).
This is the reason why the (b-i) and (c-i) as well as (b-f) and (c-f) are identical 
in the total cross section in Fig.~\ref{fig:xsec_FDvsU_l0}. 
Upon closer look at the contributions in the U gauge, these tendencies can also be seen, but 
there is no clear separation between (b) and (c) as seen in the FD gauge.

To study the interference patterns among the amplitudes in more detail, 
we now focus on the FD gauge and consider distributions at a higher energy.
Figure~\ref{fig:emuvvzh_yz_diagrams_if} shows the rapidity distribution of the $Z$ boson for $e^{-}\mu^{+}\to \nu\bar{\nu}Z_{\lambda}$ with $\lambda=0$ at $\sqrt{s}=30\ \text{TeV}$ on a linear scale this time. 
In both panels 
the black-solid lines denote the observable distribution, 
and hence they are identical.
The solid line shows that
the peak positions move to more forward and backward regions, at around $|y|=2.0$, compared with the $\sqrt{s}=1$~TeV case in Fig.~\ref{fig:dxsec_FDvsU_l0}.

In Fig.~\ref{fig:emuvvzh_yz_diagrams_if}(left), similar to Fig.~\ref{fig:dxsec_FDvsU_l0},
the contributions from the five individual FD-gauge amplitudes are shown separately.
We observe again that the VBF contribution (a) is distributed across the entire rapidity region,
and find peaks not only in the central region ($y\sim0$) but also in the  
forward and backward regions at around $|y|=3.0$ for the $\sqrt{s}=30$~TeV case.
On the other hand, we find a clear asymmetry for the (b) and (c) categories;
$Z$ bosons are emitted into the forward region via $e^-W^+$ scattering (b),
while they are emitted into the backward region via $W^-\mu^+$ scattering (c).

We point out here that, at around the center, $y\approx0$,
although the physical distribution (black solid) and the VBF contribution (red dashed)
may not appear to overlap at first glance in Fig.~\ref{fig:emuvvzh_yz_diagrams_if}(left), 
this is simply a matter of binning for making a plot and     
they actually overlap if we make the bin size finer. 
In brief, 
the total cross section is dominated by the single VBF amplitude in the FD gauge
only at the center,
while this is not the case in the U gauge as seen in Fig.~\ref{fig:dxsec_FDvsU_l0}.

In Fig.~\ref{fig:emuvvzh_yz_diagrams_if}(right), we investigate interference patterns among the FD-gauge amplitudes.
For the green dashed line, (b)=(b-i)+(b-f), for example, 
the two $Z$-emission amplitudes from the electron line in Fig.~\ref{fig:vvz} are summed before taking the absolute value squared of the amplitudes as
\begin{align}\label{eq:group_b}
  \sigma_{\rm b}\propto |{\cal M}_{{\rm b-i}}+{\cal M}_{{\rm b-f}}|^2
\end{align}
in order to see the interference between the amplitudes.
\rev{
We note that, as expected, the interference between (b) and (c) can be seen only at the very central region and very tiny in the FD gauge, namely
\begin{align}
  |{\cal M}_{{\rm b}}+{\cal M}_{{\rm c}}|^2
  \approx |{\cal M}_{{\rm b}}|^2+|{\cal M}_{{\rm c}}|^2 \ .
\end{align}
}

We find that the interference between (b-i) and (b-f) as well as between (c-i) and (c-f) is constructive 
and the contributions (b) and (c) become much larger than the observable distribution (black-solid).
As we see, only after including the VBF amplitude (a) into the sum, i.e.
the orange line with circles (a)+(b) describes the physical distribution in the forward region,
while the magenta line with crosses (a)+(c) does in the backward region.
This indicates that the interference between the VBF and the $e^-W^+/W^-\mu^+$ scattering amplitudes is destructive. 
We find that the physical distribution can be described by three FD-gauge amplitudes from the two categories, (a)+(b) or (a)+(c),
except at the very center, where only the VBF amplitude (a) dominates.

To summarize what we found in this section,
the individual U-gauge amplitudes for the longitudinally-polarized Z-boson productions have unphysical energy growth at high energies and give little useful information on the physical distributions.
In other words, the full set of the relevant Feynman amplitudes should always be taken into account in the U gauge.
In contrast, the FD-gauge amplitudes do not have such energy growth and can provide clear physics interpretation. 
In the FD gauge, for the $e^{-}\mu^{+}\to \nu\bar{\nu}Z$ process,
the Z bosons from the lepton lines tend to be emitted along the beam axis, i.e. in the forward or backward region, while
the Z bosons from the VBF are produced in the entire rapidity region.
Moreover, this allows us to examine, at the amplitude level, how each amplitude interferes with the others in different kinematic regions.
As we will see in the next section, a similar pattern of contributions and interference appears in the $e^{-}\mu^{+}\to \nu\bar{\nu}Zh$ process. The observations made in this section therefore provide a useful basis for understanding the structure of the amplitudes in that process.
We also refer to Ref.~\cite{Chen:2022xlg} for a similar study on the  $e^{-}\mu^{+}\to e^-\mu^+\gamma$ process.

We note in passing that, although we investigated the $l^{-}l^{+}\to\nu\bar{\nu}Z$ process~\eqref{ll_vvz} for a better 
understanding of the $l^{-}l^{+}\to\nu\bar{\nu}Zh$ process~\eqref{ll_vvzh},
the process~\eqref{ll_vvz} itself is also an important process 
at future lepton colliders~\cite{Hagiwara:1990gk} for the precision test of the electroweak gauge theory; see, e.g. Ref.~\cite{LinearColliderVision:2025hlt}. 
Studies in the FD gauge can be useful to extract the VBF contribution, 
which contains the triple gauge coupling.

\section{$l^-l^+\to \nu\bar{\nu}Zh$}\label{sec:vvzh}

In this section, we apply our study of the $e^-\mu^+\to \nu_{e}\bar{\nu}_{\mu}Z$ process~\eqref{vvz} in the previous section 
to the $e^-\mu^+\to \nu_{e}\bar{\nu}_{\mu}Zh$ process~\eqref{vvzh}, which is the main target in this article.
As in the previous section, we focus on the production of the longitudinally-polarized ($\lambda=0$) Z boson in the final state.

\begin{figure}
\centering
\begin{tikzpicture}[scale=1]
\begin{feynhand}
    \vertex [particle] (i1) at (-1,-2) {$e^{-}$};
    \vertex [particle] (i2) at (1,-2) {$\mu^{+}$};
    \vertex [particle] (f1) at (-1,2) {$\nu_{e}$};
    \vertex [particle] (f2) at (1,2) {$\bar{\nu}_{\mu}$};
    \vertex [NEblob, scale=0.7] (a) at (0,0) {};
    \vertex [particle] (z) at (-0.33,2) {$Z$};
    \vertex [particle] (h) at (0.33,2) {$h$};
    \vertex (w1) at (-1,0);
    \vertex (w2) at (1,0);
    \propag [fermion] (i1) to (w1);
    \propag [fermion] (w1) to (f1);
    \propag [anti fermion] (i2) to (w2);
    \propag [anti fermion] (w2) to (f2);
    \propag [charged boson] (w1) to [edge label'=$W^{-}$] (a);
    \propag [charged boson] (w2) to [edge label=$W^{+}$](a); 
    \propag [boson] (a) to (z);
    \propag [scalar] (a) to (h);
\end{feynhand}
\end{tikzpicture}
\begin{tikzpicture}[scale=1]
\begin{feynhand}
    \vertex [particle] (i1) at (-1,-2) {$e^{-}$};
    \vertex [particle] (i2) at (1,-2) {$\mu^{+}$};
    \vertex [particle] (f1) at (-1,2) {$\nu_{e}$};
    \vertex [particle] (f2) at (1,2) {$\bar{\nu}_{\mu}$};
    \vertex [NEblob, scale=0.7] (a) at (-1,0) {};
    \vertex [particle] (z) at (-0.33,2) {$Z$};
    \vertex [particle] (h) at (0.33,2) {$h$};
    \vertex (w2) at (1,0);
    \propag [fermion] (i1) to (a);
    \propag [fermion] (a) to (f1);
    \propag [anti fermion] (i2) to (w2);
    \propag [anti fermion] (w2) to (f2);
    \propag [charged boson] (w2) to [edge label=$W^{+}$](a); 
    \propag [boson] (a) to (z);
    \propag [scalar] (a) to (h);
\end{feynhand}
\end{tikzpicture}
\begin{tikzpicture}[scale=1]
\begin{feynhand}
    \vertex [particle] (i1) at (-1,-2) {$e^{-}$};
    \vertex [particle] (i2) at (1,-2) {$\mu^{+}$};
    \vertex [particle] (f1) at (-1,2) {$\nu_{e}$};
    \vertex [particle] (f2) at (1,2) {$\bar{\nu}_{\mu}$};
    \vertex [NEblob, scale=0.7] (a) at (1,0) {};
    \vertex [particle] (z) at (-0.33,2) {$Z$};
    \vertex [particle] (h) at (0.33,2) {$h$};
    \vertex (w1) at (-1,0);
    \propag [fermion] (i1) to (w1);
    \propag [fermion] (w1) to (f1);
    \propag [anti fermion] (i2) to (a);
    \propag [anti fermion] (a) to (f2);
    \propag [charged boson] (w1) to [edge label'=$W^{-}$] (a);
    \propag [boson] (a) to (z);
    \propag [scalar] (a) to (h);
\end{feynhand}
\end{tikzpicture}
(a)\hspace*{22.5mm}(b)\hspace*{22.5mm}(c)
\caption{The Feynman diagrams for $e^-\mu^+\to \nu_{e}\bar{\nu}_{\mu}Zh$ are classified into three groups; (a) vector boson scattering (VBS), (b) $e^{-}W^{+}$ scattering, and (c) $W^{-}\mu^+$ scattering.}
\label{fig:vvzh}
\end{figure}
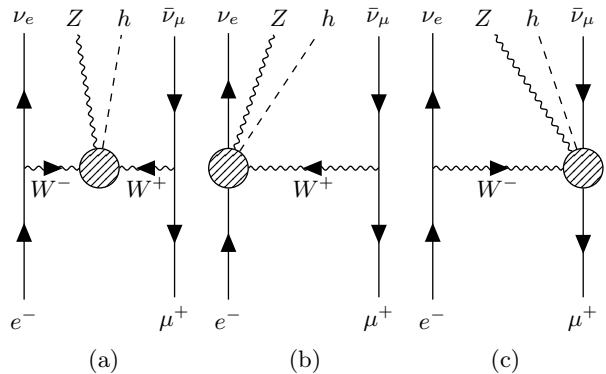

Similar to the $e^-\mu^+\to \nu_{e}\bar{\nu}_{\mu}Z$ process~\eqref{vvz}, we first classify the amplitudes of the process~\eqref{vvzh} into three main groups based on the topology of each Feynman diagram~\cite{Hagiwara:2024xdh,Hagiwara:2026lul};
\begin{enumerate}
\setlength{\itemsep}{0cm} 
 \item[(a)] vector boson scattering (VBS),
 \item[(b)] $e^-W^+$ scattering,  
 \item[(c)] $W^-\mu^+$ scattering,
\end{enumerate}
as shown in Fig.~\ref{fig:vvzh}.

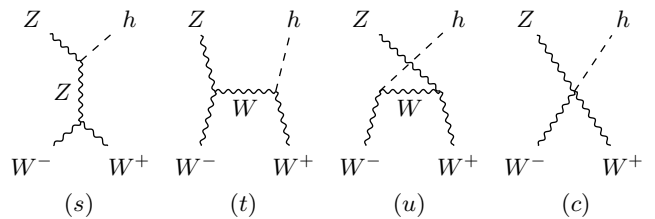
\begin{figure}
\centering
\begin{tikzpicture}[scale=0.81]
\begin{feynhand}
\vertex (a) at (0,0);
\vertex (b) at (0.8,0.7) {$h$};
\vertex (c) at (-0.8,0.7) {$Z$};
\vertex (d) at (0,-1);
\vertex (e) at (-0.8,-1.7) {$W^-$};
\vertex (f) at (0.8,-1.7) {$W^+$};
\propag [scalar] (a) to (b);
\propag [boson] (a) to (c);
\propag [boson] (a) to[edge label'=$Z$] (d);
\propag [boson] (e) to (d);
\propag [boson] (f) to (d);
\end{feynhand}
\end{tikzpicture}
\begin{tikzpicture}[scale=0.81]
\begin{feynhand}
\vertex (a) at (-0.5,-0.5);
\vertex (b) at (0.8,0.7) {$h$};
\vertex (c) at (-0.8,0.7) {$Z$};
\vertex (d) at (0.5,-0.5);
\vertex (e) at (-0.8,-1.7) {$W^-$};
\vertex (f) at (0.8,-1.7) {$W^+$};
\propag [scalar] (d) to (b);
\propag [boson] (a) to (c);
\propag [boson] (a) to[edge label'=$W$] (d);
\propag [boson] (e) to (a);
\propag [boson] (f) to (d);
\end{feynhand}
\end{tikzpicture}
\begin{tikzpicture}[scale=0.81]
\begin{feynhand}
\vertex (a) at (-0.5,-0.5);
\vertex (b) at (0.8,0.7) {$h$};
\vertex (c) at (-0.8,0.7) {$Z$};
\vertex (d) at (0.5,-0.5);
\vertex (e) at (-0.8,-1.7) {$W^-$};
\vertex (f) at (0.8,-1.7) {$W^+$};
\propag [scalar] (a) to (b);
\propag [boson] (d) to (c);
\propag [boson] (a) to[edge label'=$W$] (d);
\propag [boson] (e) to (a);
\propag [boson] (f) to (d);
\end{feynhand}
\end{tikzpicture}
\begin{tikzpicture}[scale=0.81]
\begin{feynhand}
\vertex (a) at (0,-0.5);
\vertex (b) at (0.8,0.7) {$h$};
\vertex (c) at (-0.8,0.7) {$Z$};
\vertex (e) at (-0.8,-1.7) {$W^-$};
\vertex (f) at (0.8,-1.7) {$W^+$};
\propag [scalar] (a) to (b);
\propag [boson] (a) to (c);
\propag [boson] (e) to (a);
\propag [boson] (f) to (a);
\end{feynhand}
\end{tikzpicture}
\hspace*{0mm}($s$)\hspace*{18mm}($t$)\hspace*{18mm}($u$)\hspace*{18mm}($c$)
\caption{Subdiagrams of the group (a) VBS in Fig.~\ref{fig:vvzh} in the FD gauge.}
\label{fig:fd_WW_Zh}
\end{figure}

There are four Feynman diagrams in the VBS group (a) in the FD gauge, as shown in Fig.~\ref{fig:fd_WW_Zh}; 
the $s$-channel $Z$ exchange, the $t$- and $u$-channel $W$ exchange, and the four-point contact diagrams. We note that the contact diagram is absent in the U gauge~\cite{Chen:2022gxv}.

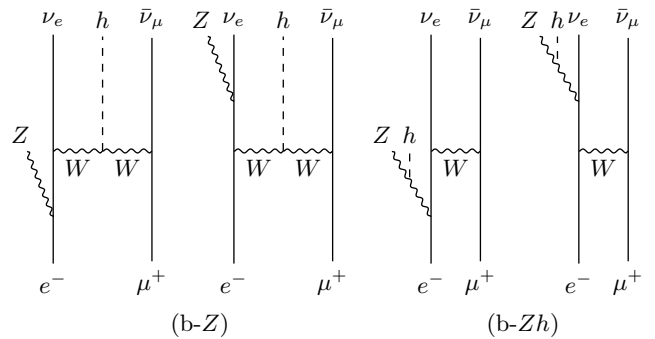
\begin{figure}
\centering
\begin{tikzpicture}[scale=0.875]
\begin{feynhand}
    \vertex [particle] (i1) at (-0.75,-2) {$e^{-}$};
    \vertex [particle] (i2) at (0.75,-2) {$\mu^{+}$};
    \vertex [particle] (f1) at (-0.75,2) {$\nu_{e}$};
    \vertex [particle] (f2) at (0.75,2) {$\bar{\nu}_{\mu}$};
    \vertex [particle] (z) at (-1.25,0.25) {$Z$};
    \vertex [particle] (h) at (0,2) {$h$};
    \vertex (a) at (0.75,0);
    \vertex (wh) at (0,0);
    \vertex (zh) at (-0.75,-1);
    \vertex (w1) at (-0.75,0);
    \propag [plain] (i1) to (w1);
    \propag [plain] (w1) to (f1);
    \propag [plain] (i2) to (a);
    \propag [plain] (a) to (f2);
    \propag [boson] (w1) to [edge label'=$W$] (wh);
    \propag [boson] (wh) to [edge label'=$W$] (a);
    \propag [scalar] (wh) to (h);
    \propag [boson] (zh) to (z);
\end{feynhand}
\end{tikzpicture}
\begin{tikzpicture}[scale=0.875]
\begin{feynhand}
    \vertex [particle] (i1) at (-0.75,-2) {$e^{-}$};
    \vertex [particle] (i2) at (0.75,-2) {$\mu^{+}$};
    \vertex [particle] (f1) at (-0.75,2) {$\nu_{e}$};
    \vertex [particle] (f2) at (0.75,2) {$\bar{\nu}_{\mu}$};
    \vertex [particle] (z) at (-1.25,2) {$Z$};
    \vertex [particle] (h) at (0,2) {$h$};
    \vertex (a) at (0.75,0);
    \vertex (wh) at (0,0);
    \vertex (zh) at (-0.75,0.75);
    \vertex (w1) at (-0.75,0);
    \propag [plain] (i1) to (w1);
    \propag [plain] (w1) to (f1);
    \propag [plain] (i2) to (a);
    \propag [plain] (a) to (f2);
    \propag [boson] (w1) to [edge label'=$W$] (wh);
    \propag [boson] (wh) to [edge label'=$W$] (a);
    \propag [scalar] (wh) to (h);
    \propag [boson] (zh) to (z);
\end{feynhand}
\end{tikzpicture}
\begin{tikzpicture}[scale=0.875]
\begin{feynhand}
    \vertex [particle] (i1) at (-0.375,-2) {$e^{-}$};
    \vertex [particle] (i2) at (0.375,-2) {$\mu^{+}$};
    \vertex [particle] (f1) at (-0.375,2) {$\nu_{e}$};
    \vertex [particle] (f2) at (0.375,2) {$\bar{\nu}_{\mu}$};
    \vertex [particle] (z) at (-1.125,0.25) {$Z$};
    \vertex [particle] (h) at (-0.7,0.25) {$h$};
    \vertex (a) at (0.375,0);
    \vertex (wh) at (0,0);
    \vertex (zh) at (-0.375,-1);
    \vertex (w1) at (-0.375,0);
	\vertex (hzz) at (-0.7,-0.4);
    \propag [plain] (i1) to (w1);
    \propag [plain] (w1) to (f1);
    \propag [plain] (i2) to (a);
    \propag [plain] (a) to (f2);
    \propag [boson] (w1) to [edge label'=$W$] (a);
    \propag [scalar] (hzz) to (h);
    \propag [boson] (zh) to (z);
\end{feynhand}
\end{tikzpicture}
\begin{tikzpicture}[scale=0.875]
\begin{feynhand}
    \vertex [particle] (i1) at (-0.375,-2) {$e^{-}$};
    \vertex [particle] (i2) at (0.375,-2) {$\mu^{+}$};
    \vertex [particle] (f1) at (-0.375,2) {$\nu_{e}$};
    \vertex [particle] (f2) at (0.375,2) {$\bar{\nu}_{\mu}$};
    \vertex [particle] (z) at (-1.125,2) {$Z$};
    \vertex [particle] (h) at (-0.7,2) {$h$};
    \vertex (a) at (0.375,0);
    \vertex (wh) at (0,0);
    \vertex (zh) at (-0.375,0.75);
    \vertex (w1) at (-0.375,0);
	\vertex (hzz) at (-0.7,1.4);
    \propag [plain] (i1) to (w1);
    \propag [plain] (w1) to (f1);
    \propag [plain] (i2) to (a);
    \propag [plain] (a) to (f2);
    \propag [boson] (w1) to [edge label'=$W$] (a);
    \propag [scalar] (hzz) to (h);
    \propag [boson] (zh) to (z);
\end{feynhand}
\end{tikzpicture}
\hspace*{10mm}(b-$Z$) \hspace*{32mm} (b-$Zh$)
\caption{The group (b) $e^-W^+$ scattering in Fig.~\ref{fig:vvzh} is further classified into two subgroups; (b-$Z$) $Z$ emissions and (b-$Zh$) $Zh$ emissions from the electron line.}
\label{fig:group_b}
\end{figure}

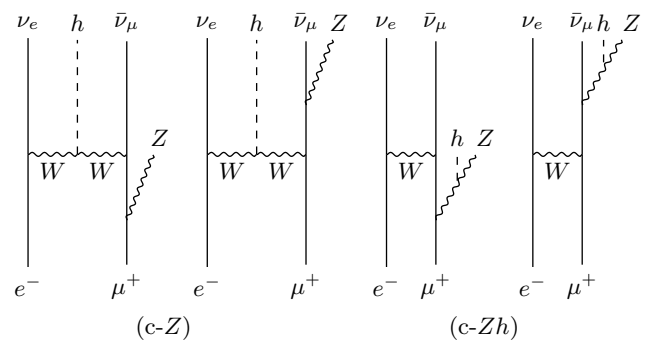
\begin{figure}
\centering
\begin{tikzpicture}[scale=0.875]
\begin{feynhand}
    \vertex [particle] (i1) at (-0.75,-2) {$e^{-}$};
    \vertex [particle] (i2) at (0.75,-2) {$\mu^{+}$};
    \vertex [particle] (f1) at (-0.75,2) {$\nu_{e}$};
    \vertex [particle] (f2) at (0.75,2) {$\bar{\nu}_{\mu}$};
    \vertex [particle] (z) at (1.25,0.25) {$Z$};
    \vertex [particle] (h) at (0,2) {$h$};
    \vertex (a) at (0.75,0);
    \vertex (wh) at (0,0);
    \vertex (zh) at (0.75,-1);
    \vertex (w1) at (-0.75,0);
    \propag [plain] (i1) to (w1);
    \propag [plain] (w1) to (f1);
    \propag [plain] (i2) to (a);
    \propag [plain] (a) to (f2);
    \propag [boson] (w1) to [edge label'=$W$] (wh);
    \propag [boson] (wh) to [edge label'=$W$] (a);
    \propag [scalar] (wh) to (h);
    \propag [boson] (zh) to (z);
\end{feynhand}
\end{tikzpicture}
\begin{tikzpicture}[scale=0.875]
\begin{feynhand}
    \vertex [particle] (i1) at (-0.75,-2) {$e^{-}$};
    \vertex [particle] (i2) at (0.75,-2) {$\mu^{+}$};
    \vertex [particle] (f1) at (-0.75,2) {$\nu_{e}$};
    \vertex [particle] (f2) at (0.75,2) {$\bar{\nu}_{\mu}$};
    \vertex [particle] (z) at (1.25,2) {$Z$};
    \vertex [particle] (h) at (0,2) {$h$};
    \vertex (a) at (0.75,0);
    \vertex (wh) at (0,0);
    \vertex (zh) at (0.75,0.75);
    \vertex (w1) at (-0.75,0);
    \propag [plain] (i1) to (w1);
    \propag [plain] (w1) to (f1);
    \propag [plain] (i2) to (a);
    \propag [plain] (a) to (f2);
    \propag [boson] (w1) to [edge label'=$W$] (wh);
    \propag [boson] (wh) to [edge label'=$W$] (a);
    \propag [scalar] (wh) to (h);
    \propag [boson] (zh) to (z);
\end{feynhand}
\end{tikzpicture}
\begin{tikzpicture}[scale=0.875]
\begin{feynhand}
    \vertex [particle] (i1) at (-0.375,-2) {$e^{-}$};
    \vertex [particle] (i2) at (0.375,-2) {$\mu^{+}$};
    \vertex [particle] (f1) at (-0.375,2) {$\nu_{e}$};
    \vertex [particle] (f2) at (0.375,2) {$\bar{\nu}_{\mu}$};
    \vertex [particle] (z) at (1.125,0.25) {$Z$};
    \vertex [particle] (h) at (0.7,0.25) {$h$};
    \vertex (a) at (0.375,0);
    \vertex (wh) at (0,0);
    \vertex (zh) at (0.375,-1);
    \vertex (w1) at (-0.375,0);
	\vertex (hzz) at (0.7,-0.4);
    \propag [plain] (i1) to (w1);
    \propag [plain] (w1) to (f1);
    \propag [plain] (i2) to (a);
    \propag [plain] (a) to (f2);
    \propag [boson] (w1) to [edge label'=$W$] (a);
    \propag [scalar] (hzz) to (h);
    \propag [boson] (zh) to (z);
\end{feynhand}
\end{tikzpicture}
\begin{tikzpicture}[scale=0.875]
\begin{feynhand}
    \vertex [particle] (i1) at (-0.375,-2) {$e^{-}$};
    \vertex [particle] (i2) at (0.375,-2) {$\mu^{+}$};
    \vertex [particle] (f1) at (-0.375,2) {$\nu_{e}$};
    \vertex [particle] (f2) at (0.375,2) {$\bar{\nu}_{\mu}$};
    \vertex [particle] (z) at (1.125,2) {$Z$};
    \vertex [particle] (h) at (0.7,2) {$h$};
    \vertex (a) at (0.375,0);
    \vertex (wh) at (0,0);
    \vertex (zh) at (0.375,0.75);
    \vertex (w1) at (-0.375,0);
	\vertex (hzz) at (0.7,1.4);
    \propag [plain] (i1) to (w1);
    \propag [plain] (w1) to (f1);
    \propag [plain] (i2) to (a);
    \propag [plain] (a) to (f2);
    \propag [boson] (w1) to [edge label'=$W$] (a);
    \propag [scalar] (hzz) to (h);
    \propag [boson] (zh) to (z);
\end{feynhand}
\end{tikzpicture}
\hspace*{0mm}(c-$Z$) \hspace*{32mm} (c-$Zh$)
\caption{The group (c) $W^-\mu^+$ scattering in Fig.~\ref{fig:vvzh} is further classified into two subgroups; (c-$Z$) $Z$ emissions and (c-$Zh$) $Zh$ emissions from the muon line.}
\label{fig:group_c}
\end{figure}

The $e^-W^+$ group (b) is further classified into two subgroups; 
\begin{enumerate}[label=(b-\roman*), align=left, leftmargin=2.5em]
	\setlength{\itemsep}{0cm} 
 \item[ (b-$Z$)] $Z$ emission from the VBF Higgs-production diagram,
 \item[ (b-$Zh$)] $Zh$ emission from the $e^-\mu^+\to \nu_e \bar{\nu}_\mu$ diagram,
\end{enumerate}
as shown in Fig.~\ref{fig:group_b}. 
Both subgroups each have two diagrams; $Z$ or $Zh$ emissions from the initial and the final electron line.
The same applies to the $W^-\mu^+$ group (c), but from the muon line as shown in Fig.~\ref{fig:group_c}.
In this section $Z$ or $Zh$ emissions from the initial and final states are considered as a single group.

To sum up, there are twelve (eleven) Feynman diagrams in total in the FD (U) gauge
for the process~\eqref{vvzh}. 
As in the previous section, we study the contributions from each group to the total and differential cross sections
both in the FD and U gauges.
As mentioned in Sec.~\ref{sec:intro},
the process includes the Higgs interaction with W and Z bosons simultaneously,
and hence it is sensitive to physics related to electroweak gauge theory.

\begin{figure}
  \includegraphics[width=1\columnwidth]{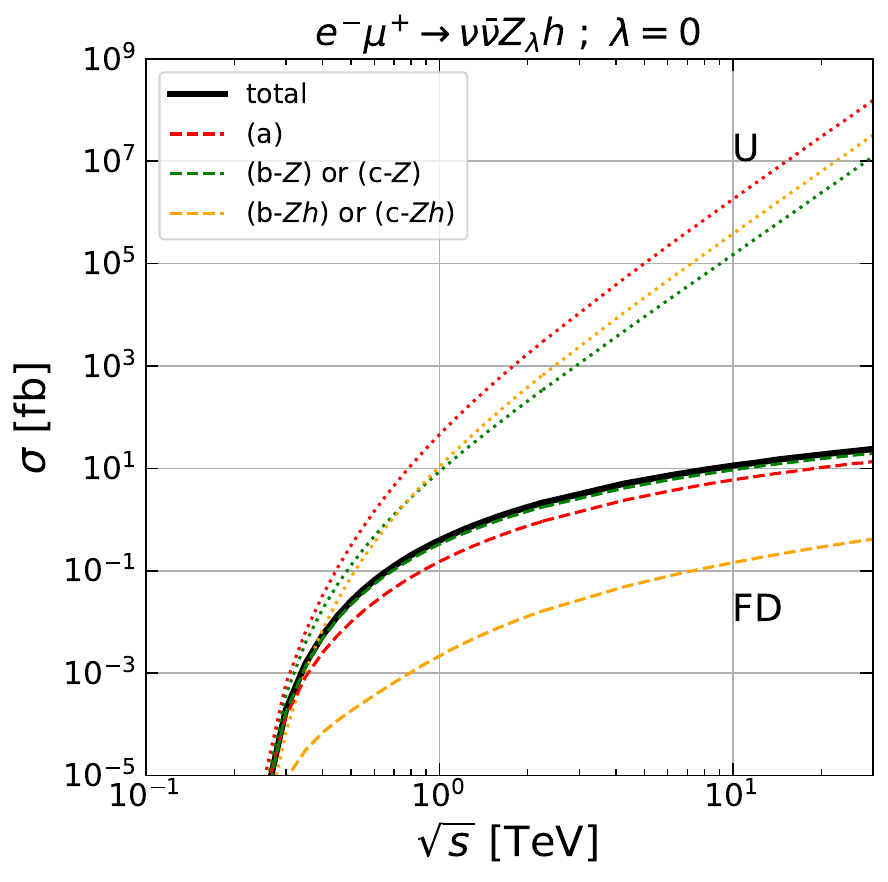}
  \caption{Total cross section of $e^{-}\mu^{+}\to \nu\bar{\nu}Z_{\lambda}h$ for $\lambda=0$
as a function of the collision energy,
where $\lambda$ is the helicity of the Z boson in the $e^-\mu^+$ collision c.m. frame.
The solid line denotes the total cross section, while dashed (dotted) lines show contributions from each diagram category defined in Figs.~\ref{fig:vvzh}--\ref{fig:group_c} in the FD (U) gauge.}
\label{fig:vvzh_xsec_FDvsU_l0}
\end{figure}

Figure~\ref{fig:vvzh_xsec_FDvsU_l0} shows the total cross section of $e^-\mu^+\to \nu_e\bar{\nu}_\mu Zh$ for $\lambda=0$ as a function of the collision energy $\sqrt{s}$ from 100~GeV up to 30~TeV, where $\lambda$ is the helicity of the final-state $Z$ boson in the $e^-\mu^+$ collision c.m. frame.
The black-solid line, which is identical to the red-dashed line for $e^-\mu^+\to \nu_e\bar{\nu}_\mu Zh$ in Fig.~\ref{fig:xsec},
denotes the physical cross section, and does not depend on the gauge choice.
The contributions from each group are shown by the dashed (dotted) lines in the FD (U) gauge.
The red, orange, and green lines show the contributions from (a) VBS, 
(b-$Z$) or (c-$Z$) $Z$ emissions from the VBF Higgs-production diagram, and (b-$Zh$) or (c-$Zh$) $Zh$ emissions from the $e^-\mu^+\to \nu_e \bar{\nu}_\mu$ diagram, respectively. 
The lines of (b-$Z$) and (c-$Z$) as well as (b-$Zh$) and (c-$Zh$) completely overlap.

In the U gauge, shown by the dotted lines, similar to the $\nu\bar{\nu}Z$ case in Fig.~\ref{fig:xsec_FDvsU_l0}, 
each contribution rapidly grows with energy even with a higher power as $(\sqrt{s})^4$ at high energies.
This implies that significant gauge cancellation among the amplitudes is required 
to obtain the physical cross section. 
We point out here that the naive opposite-sign hypothesis between the $hWW$ and $hZZ$ couplings violates gauge invariance 
for the VBS $Vh$ production, giving rise not only to 
preventing the cancellation among the amplitudes, but actually to largely enhancing the cross section in the U gauge.
As a result, the severe constraints on the relative sign of those couplings are already obtained at the LHC~\cite{ATLAS:2024vxc,CMS:2024srp} as mentioned in Sec.~\ref{sec:intro}.

In the FD gauge, denoted by the dashed lines, in contrast, the contributions from the VBS (red-dashed) and the $Z$ emission (green-dashed) are along with the total cross section without any unphysical energy growth.
The contributions from the $Zh$ emission (orange-dashed) are about two orders of magnitude smaller than the others.
This agrees with our naive expectation that
the emission of a $Zh$ pair from the fermion line 
should be suppressed more significantly than the single $Z$ emission.
Such a hierarchy among the contributions from each group is not observed in the U gauge.  
Different from Fig.~\ref{fig:xsec_FDvsU_l0}, 
the $Z$ or $Zh$ emissions from the initial and final states are grouped in Fig.~\ref{fig:vvzh_xsec_FDvsU_l0}, e.g.
the two $Z$-emission amplitudes from the electron line in (b-$Z$), depicted in Fig.~\ref{fig:group_b}, 
are summed before taking the absolute value squared
as in Eq.~\eqref{eq:group_b}.
If we look into the individual contributions, 
we find similar cross-section ratios in Eq.~\eqref{eq:ratio} between the initial and final state radiations both in the FD and U gauges. 

We note here that, different from that in the $l^-l^+\to \nu\bar{\nu}t\bar th$ process~\cite{Hagiwara:2024xdh,Hagiwara:2026lul}, 
the VBS contribution in the FD gauge for this process does not dominate even at high energies,
and the contributions from the emission of the $Z$ boson from the fermion line are also significant.

\begin{figure}
  \includegraphics[width=1\columnwidth]{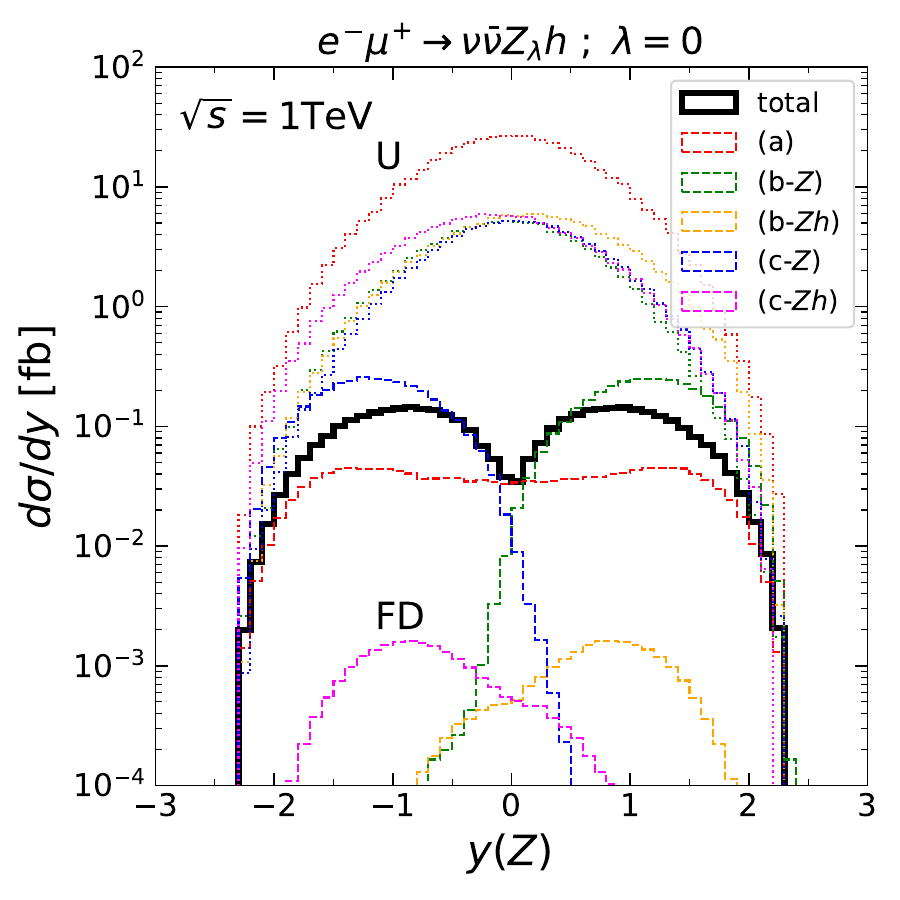}
  \caption{Rapidity distribution of the Z boson for $e^{-}\mu^{+}\to \nu\bar{\nu}Z_{\lambda}h$ with $\lambda=0$ at $\sqrt{s}=1$~TeV,
where $\lambda$ is the helicity of the Z boson in the $e^-\mu^+$ collision c.m. frame. 
The solid line denotes the observable distribution, while dashed (dotted) lines show contributions from each diagram category defined in Figs.~\ref{fig:vvzh}--\ref{fig:group_c} in the FD (U) gauge.
}
\label{fig:vvzh_yz_FDvsU_l0}
\end{figure}

Let us move to kinematical distributions for $e^{-}\mu^{+}\to \nu\bar{\nu}Zh$.
The rapidity distribution of the Z boson for $\lambda=0$ at $\sqrt{s}=1\ \text{TeV}$ is shown in Fig.~\ref{fig:vvzh_yz_FDvsU_l0}.
The line styles are the same as those in Fig.~\ref{fig:vvzh_xsec_FDvsU_l0}, 
but the degeneracies between (b-$Z$) and (c-$Z$) as well as between (b-$Zh$) and (c-$Zh$) are resolved.
What we find here is very similar to that in the $\nu\bar\nu Z$ case discussed in Fig.~\ref{fig:dxsec_FDvsU_l0}.
\rev{
See also the transverse-momentum distributions in Appendix. 
}

The black-solid line, i.e. the observable distribution, shows that the central Z-boson production ($y\sim0$) is suppressed and the distribution has  
peaks at around $|y|=0.8$.

In the U gauge (dotted lines), again, 
not only the magnitude
but also the shapes of the distributions from the individual groups are completely different from the observable distribution.

In the FD gauge (dashed lines), in contrast, the properties of the Feynman diagrams for each group defined in Fig.~\ref{fig:vvzh} are apparently reflected in the distributions. 
As shown in Fig.~\ref{fig:fd_WW_Zh}, the group (a) VBS, i.e. the $W^-W^+\to Zh$ scattering, has the four ($s$-, $t$-, $u$- and $c$-channel) amplitudes in the FD gauge, 
and gives an isotropic rapidity distribution for the longitudinally-polarized Z boson,
 denoted by the red-dashed line.
On the other hand, the Z bosons from the group (b) $e^-W^+$ scattering in Fig.~\ref{fig:group_b} and from the group (c) $W^-\mu^+$ scattering in Fig.~\ref{fig:group_c}
show strong asymmetry in the $y(Z)$ distribution.
For the $e^-W^+$ scattering (b), the Z bosons are emitted along the electron line,
and hence they tend to be scattered into the forward ($y>0$) region.
The green and orange dashed lines are consistent with this expectation.
Similarly, for the $W^-\mu^+$ scattering (c),
the Z bosons are emitted along the muon line,
and hence they tend to be scattered into the backward ($y<0$) region.
The blue and magenta dashed lines are consistent with this expectation.
The two contributions from $e^-W^+$ and $W^-\mu^+$ scatterings are symmetric with each other under $y \leftrightarrow -y$.
As seen in the total cross section in Fig.~\ref{fig:vvzh_xsec_FDvsU_l0}, 
the contributions from the $Zh$ emissions (b-$Zh$) and (c-$Zh$) are 
about two orders of magnitude smaller than those from the $Z$ emissions (b-$Z$) and (c-$Z$).	

Similar to Fig.~\ref{fig:emuvvzh_yz_diagrams_if} for the $\nu\bar\nu Z$ case,
we now study the interference patterns among the diagram categories in the FD gauge for the $\nu\bar\nu Zh$ case in more detail.
We discuss the distributions of the Higgs boson in the end.

\begin{figure*}
    \includegraphics[width=0.495\textwidth]{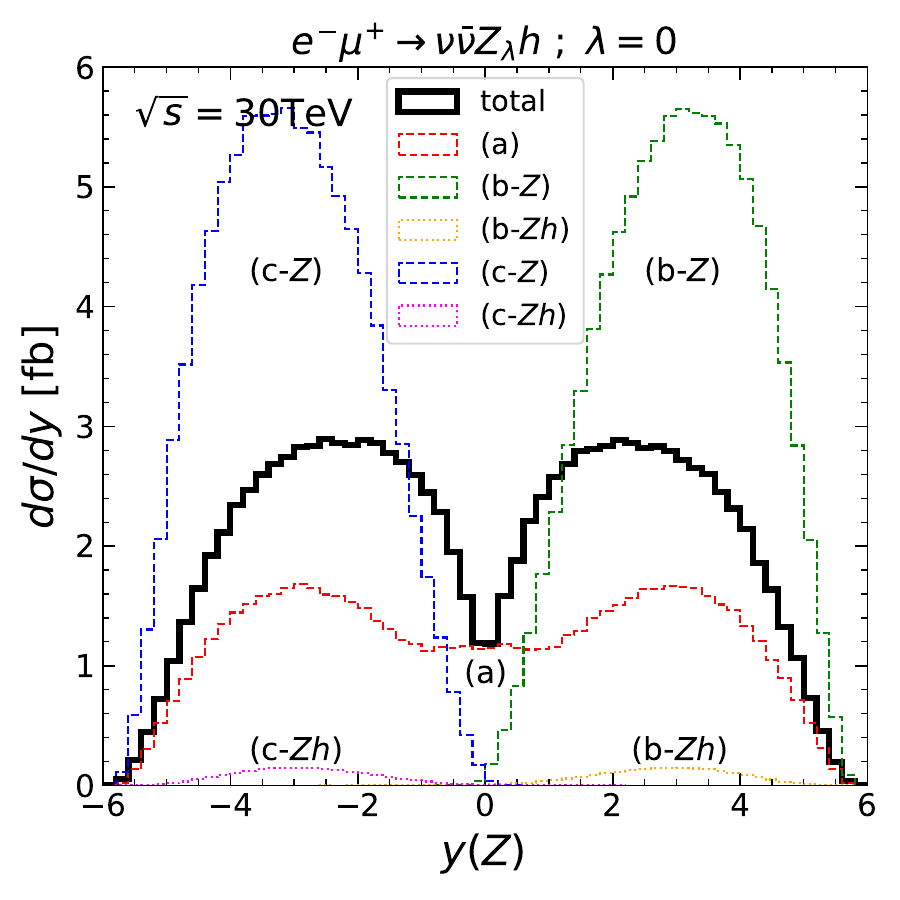}
    \includegraphics[width=0.495\textwidth]{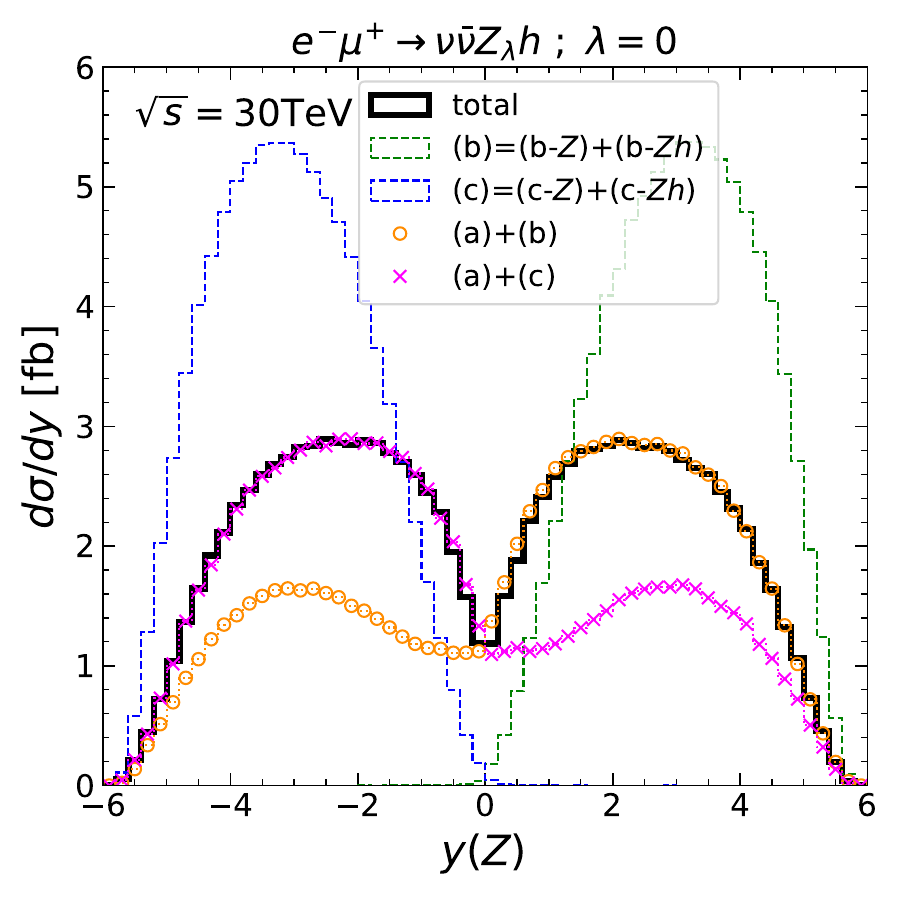}\\
    \includegraphics[width=0.495\textwidth]{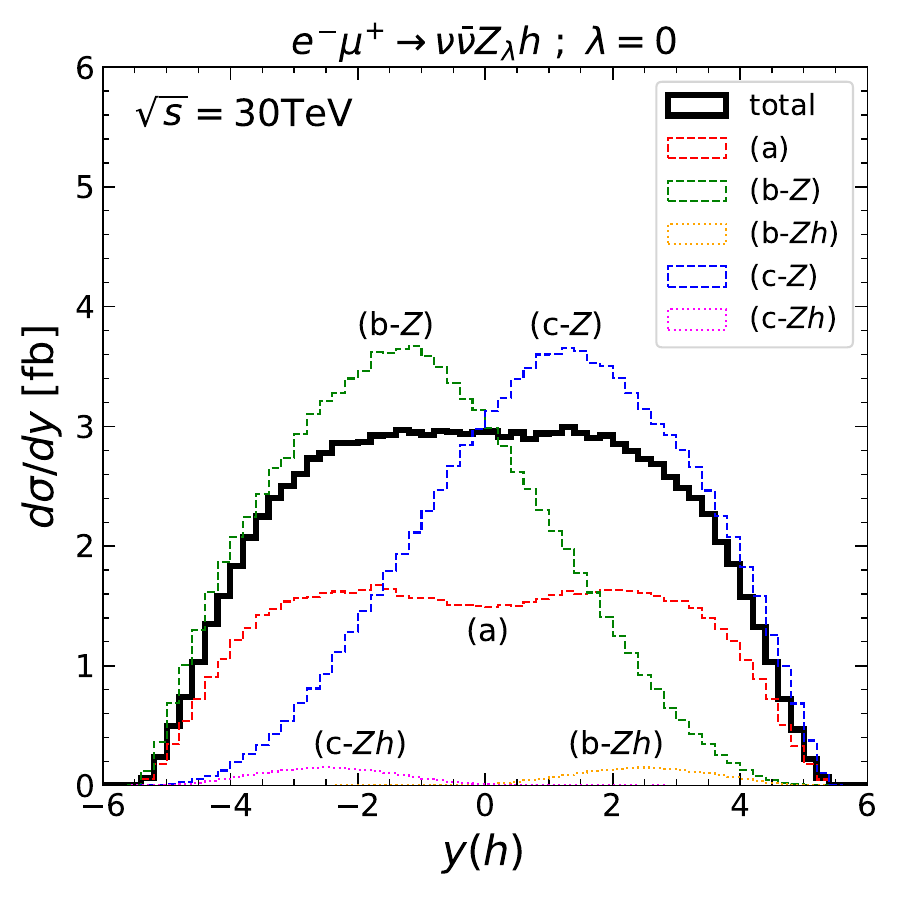}
    \includegraphics[width=0.495\textwidth]{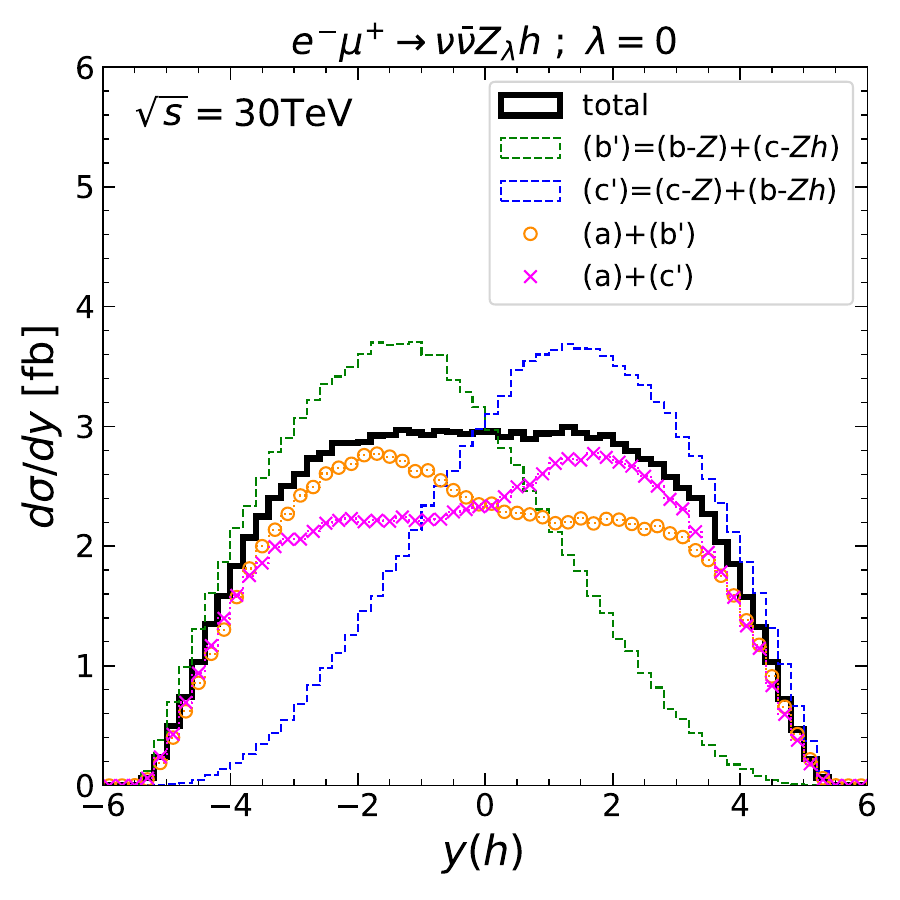}
    \caption{Rapidity distributions of the Z boson (top) and the Higgs boson (bottom) 
    for $e^{-}\mu^{+}\to \nu\bar{\nu}Z_{\lambda}h$ with $\lambda = 0$ at $\sqrt{s}=30$~TeV, where the observable distributions are shown by black-solid lines.
    The left panels show the contributions from the five individual groups defined in Figs.~\ref{fig:vvzh}-\ref{fig:group_c} in the FD gauge, while the right panels show the contributions from the sum of the two or three groups.
	}
    \label{fig:emuvvz0h_yz_yh_30TeV_if}
\end{figure*}

The top panels in Fig.~\ref{fig:emuvvz0h_yz_yh_30TeV_if} show the rapidity distribution of the $Z$ boson for $e^{-}\mu^{+}\to \nu\bar{\nu}Z_{\lambda}h$ with $\lambda=0$ at $\sqrt{s}=30$~TeV on a linear scale. 
In both of the left and right panels 
the black-solid lines denote the observable distribution, 
and hence they are identical.
The solid line shows that
the peak positions move to more forward and backward regions, at around $|y(Z)|=2.0$, compared with the $\sqrt{s}=1$~TeV case in Fig.~\ref{fig:vvzh_yz_FDvsU_l0}.

In Fig.~\ref{fig:emuvvz0h_yz_yh_30TeV_if}(top-left), similar to Fig.~\ref{fig:vvzh_yz_FDvsU_l0},
the contributions from the five individual groups are shown separately only in the FD gauge.
Due to the increase of the collision energy from 1~TeV to 30~TeV,
the Z bosons via the VBS (a) tend to be produced a bit more in the forward and backward region, peaked at around $|y(Z)|=3.0$,
while we find a clear asymmetry for the (b) and (c) categories;
$Z$ bosons are emitted into the forward region via $e^-W^+$ scattering (b),
while they are emitted into the backward region via $W^-\mu^+$ scattering (c).
The physical distribution is dominated by the VBS contribution 
only at the center, $y\approx0$.
The contributions from (b-$Z$) and (c-$Z$) are much larger than the 
observable distribution because of the constructive interference between the 
initial and final $Z$-emission amplitudes,
which can also be seen in Fig.~\ref{fig:emuvvzh_yz_diagrams_if}(right) for the $\nu\bar\nu Z$ case.
The contributions from (b-$Zh$) and (c-$Zh$) are small, but not negligible.

In Fig.~\ref{fig:emuvvz0h_yz_yh_30TeV_if}(top-right), 
we examine interference patterns among the subgroups.
The green-dashed line (b) in the right panel is slightly smaller than (b-$Z$) in the left panel,
indicating a destructive interference between (b-$Z$) and (b-$Zh$) in the forward region. 
The same destructive interference is observed between (c-$Z$) and (c-$Zh$), shown by the blue-dashed line (c), in the backward region. 
Similar to Fig.~\ref{fig:emuvvzh_yz_diagrams_if}(right),
only after including the VBS amplitudes (a) into the sum, i.e.
the orange line with circles (a)+(b) describes the physical distribution in the forward region,
while the magenta line with crossed (a)+(c) does in the backward region.
This indicates that the interference between the VBS and the $e^-W^+/W^-\mu^+$ scattering amplitudes is destructive. 

We now turn to the distribution of the Higgs boson. 
The bottom-left panel in Fig.~\ref{fig:emuvvz0h_yz_yh_30TeV_if} shows the rapidity distribution of the Higgs boson. 
The physical $y(h)$ distribution, shown by the black-solid line, is isotropic and rather different from the $y(Z)$ distribution,
contrary to our naive expectation for the productions with the longitudinally-polarized Z bosons. 
This difference between the $y(h)$ and $y(Z)$ distributions can be understood by the difference of the contributions from each group as explained below.

Firstly, the contribution from the VBS (a), denoted by the red-dashed line, is similar 
in $y(h)$ and $y(Z)$.
Since the longitudinally-polarized ($\lambda=0$) Z bosons are considered here,
both of the Higgs bosons and the Z bosons from the 
$W^-W^+$ scattering tend to be produced isotropically. 
We note that, unlike the $y(Z)$ distribution, the VBS contribution in $y(h)$ never dominates 
even at the center, $y\approx0$. 

Secondly, we find that the (b-$Z$) and (c-$Z$) contributions, denoted by the green-dashed line and the blue-dashed line, respectively, 
 for the Higgs boson are swapped
relative to the Z-boson distribution, 
i.e. (b-$Z$) in the backward region and (c-$Z$) in the forward region in $y(h)$,
although the distribution of each of these groups becomes broader and they overlap significantly in the 
central region. 
The reason is as follows.
Without a $Z$ emission, i.e. the Higgs bosons via the VBF ($W^-W^+\to h$) are produced 
isotropically. 
Due to a $Z$ emission from the fermion line, the $W^-W^+\to h$ system tends to be boosted
into the opposite direction to the Z boson,
although the boost effect is not strong enough to completely separate the contributions from (b-$Z$) and (c-$Z$).
 
Finally, the contributions from (b-$Zh$) and (c-$Zh$), denoted by the orange-dotted line and the magenta-dotted line, respectively, are very similar in $y(h)$ and $y(Z)$.
This is because both of the Higgs boson and the Z boson from (b-$Zh$) and (c-$Zh$) are emitted together along the electron line and the muon line, respectively.

As seen, all the behaviors from each category in the FD gauge agree with our naive expectation based on the topology of the Feynman diagrams in Figs.~\ref{fig:vvzh}-\ref{fig:group_c}.
We find that the difference of the observable distributions for $y(h)$ and $y(Z)$ in Fig.~\ref{fig:emuvvz0h_yz_yh_30TeV_if} mainly 
stems from the different distributions for the Higgs boson and the Z boson
from the (b-$Z$) and (c-$Z$) subgroups.
It should be noted that, in the U gauge, 
the distributions from the individual groups for the Higgs boson are very similar to the distributions for the Z boson; 
see Fig.~\ref{fig:vvzh_yz_FDvsU_l0} for the $y(Z)$ distributions at $\sqrt{s}=1$~TeV. 
Therefore, they provide no useful information to understand the observable distributions.

In Fig.~\ref{fig:emuvvz0h_yz_yh_30TeV_if}(bottom-right),
we now examine interference patterns among the groups in $y(h)$.
Since the positions of (b-$Z$) and (c-$Z$) are swapped 
relative to the $y(Z)$ distribution, 
we take different combinations;
(b')=(b-$Z$)+(c-$Zh$) and 
(c')=(c-$Z$)+(b-$Zh$).
The green-dashed line (b') in the right panel is very similar to 
(b-$Z$) in the left panel.
If we look closely, (b') is slightly larger than (b-$Z$) at around the peak position.
We find that the interference between (b-$Z$) and (c-$Zh$) is very small,
and the (b') contribution is almost equal to the sum of the contributions 
from the absolute value squared of the amplitudes in each group as
\begin{align}
  \sigma_{\rm b'}\propto |{\cal M}_{{\rm b-}Z}+{\cal M}_{{\rm c-}Zh}|^2
  \approx |{\cal M}_{{\rm b-}Z}|^2+|{\cal M}_{{\rm c-}Zh}|^2 \ .
\end{align}
The same can be applied for (c'), denoted by the blue-dashed line.

Similar to the $y(Z)$ distribution,
the orange line with circles (a)+(b') indicates a destructive interference between the VBS contribution (a) and (b').
The same destructive interference between (a) and (c') can be seen by the magenta line with crosses. 
However, unlike the $y(Z)$ distribution, 
since (b') and (c') significantly overlap each other,
even after taking into account the VBS contribution,
(a)+(b') cannot fully describe the observable distribution for $y(h)<0$, especially in the central region.
Similarly, (a)+(c') does not fully agree with the physical distribution for $y(h)>0$.
In other words, the full set of the amplitudes should be taken into account for the $y(h)$ distribution.
We note that, if we take (b)=(b-$Z$)+(b-$Zh$) ((c)=(c-$Z$)+(c-$Zh$)) as in $y(Z)$, 
the agreement between the black-solid line and the orange-circle (magenta-cross) line in the backward (forward) region
becomes even worse.

\begin{figure}
	\center
    \includegraphics[width=1\columnwidth]{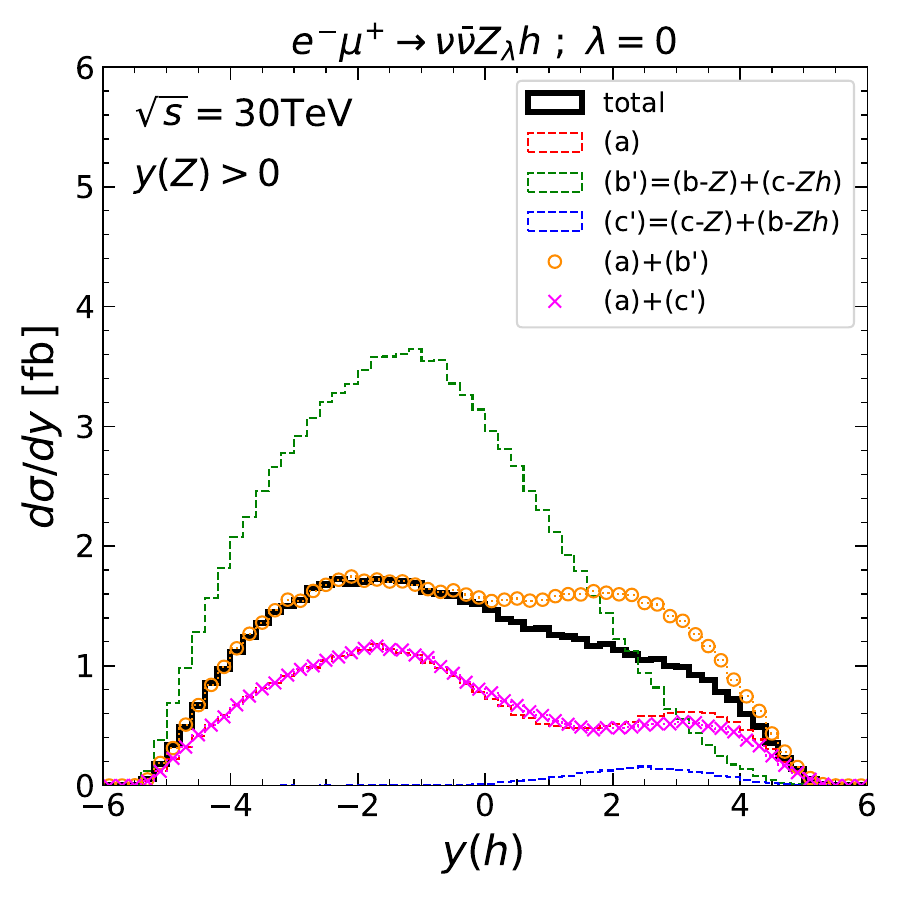}
    \caption{
Rapidity distributions of the Higgs boson
    for $e^{-}\mu^{+}\to \nu\bar{\nu}Z_{\lambda}h$ with $\lambda = 0$ at $\sqrt{s}=30$~TeV, where the $y(Z) > 0$ cut is imposed.
The line styles are the same as those in Fig.~\ref{fig:emuvvz0h_yz_yh_30TeV_if}.
	}
    \label{fig:emuvvz0h_yz_yh_30TeV_if_yzcut0}
\end{figure}

The situation can be improved by some kinematical cuts as discussed below.
Similar to the bottom-right panel in Fig.~\ref{fig:emuvvz0h_yz_yh_30TeV_if},
we show the $y(h)$ distribution in
Fig.~\ref{fig:emuvvz0h_yz_yh_30TeV_if_yzcut0}, where we 
impose the kinematical cut
\begin{align} \label{yzcut}
	y(Z)>0\ ,
\end{align}
motivated from the top panels in Fig.~\ref{fig:emuvvz0h_yz_yh_30TeV_if}
in order to remove the contributions from $W^-\mu^+$ scattering (c).
With the cut in Eq.~\eqref{yzcut}, we obtain (b')$\approx$(b-$Z$) and (c')$\approx$(b-$Zh$) as expected. 

Now we find that the observable $y(h)$ distribution (black-solid) in the backward region is completely dominated by (a)+(b')$\approx$(a)+(b-$Z$), denoted by the orange line with circles.
Conversely, if we take the $y(Z)<0$ cut, (a)+(c')$\approx$(a)+(c-$Z$), the magenta line with crosses, can describe the physical distribution in the forward region.

This demonstrates that FD-gauge amplitudes allow us to identify which contributions are relevant in specific kinematic regions and 
can provide a clear way to visualize appropriate kinematic cuts 
to enhance or remove specific contributions 
for processes under consideration.

{\section{Summary}\label{sec:summary}

In this paper, 
\rev{
we studied the $l^-l^+\to \nu\bar{\nu}Zh$ process, which simultaneously involves both the $hWW$ and $hZZ$ interactions and hence is sensitive to the electroweak symmetry-braking sector, for future TeV-scale lepton colliders.
In order to understand the kinematical distributions for the Z boson in the $\nu\bar{\nu}Zh$ final state, 
we also considered the $l^-l^+\to \nu\bar{\nu}Z$ process in detail. 
In this summary, we consider the $\nu\bar{\nu}Z$ and $\nu\bar{\nu}Zh$ final states in parallel and discuss them together.
}

We classified the amplitudes into three main groups based on the topology of each Feynman diagram depicted in Figs.~\ref{fig:vvz} and \ref{fig:vvzh};
(a) vector boson fusion/scattering (VBF/VBS), (b) $l^-W^+$ scattering, and (c) $W^-l^+$ scattering,
and studied the interference patterns among the amplitudes both in the recently proposed Feynman-diagram (FD) gauge~\cite{Hagiwara:2020tbx,Chen:2022gxv,Chen:2022xlg} and in the unitary (U) gauge.  

Focusing on the productions for longitudinally-polarized Z bosons in this paper,
we showed in Figs.~\ref{fig:xsec_FDvsU_l0} and \ref{fig:vvzh_xsec_FDvsU_l0} that subtle gauge cancellation among the amplitudes at high energies in the U gauge is absent in the FD gauge for both processes.
In addition, in Figs.~\ref{fig:dxsec_FDvsU_l0} and \ref{fig:vvzh_yz_FDvsU_l0},
we found that the U-gauge amplitudes give little useful information on the physical distributions,
while the FD-gauge amplitudes can provide clear physics interpretation. 

In Figs.~\ref{fig:emuvvzh_yz_diagrams_if} and \ref{fig:emuvvz0h_yz_yh_30TeV_if},
we investigated interference patterns among the FD-gauge amplitudes in detail. 
We found that the rapidity distribution of the Z boson and the interference patterns among the amplitudes for the $l^-l^+\to \nu\bar{\nu}Zh$ process are very similar to those for the $l^-l^+\to \nu\bar{\nu}Z$ process,
and identified sizable destructive interference effects between the VBS and $l^-W^+/W^-l^+$ subamplitudes.

We also found that
the rapidity distribution of the Higgs boson for $l^-l^+\to \nu\bar{\nu}Zh$ is rather different from that of the Z boson in Fig.~\ref{fig:emuvvz0h_yz_yh_30TeV_if};
the Higgs bosons are produced isotropically, while the longitudinally-polarized Z bosons tend to be produced in the forward and backward regions.
Thanks to the property of FD-gauge amplitudes that topologies of Feynman diagrams are reflected in distributions,
we could understand that 
the difference between the observable distributions for $y(h)$ and $y(Z)$ mainly 
stems from the different distributions for the Higgs boson and the Z boson
from the VBF Higgs-production diagram, i.e. the (b-$Z$) and (c-$Z$) subgroups defined in Figs.~\ref{fig:group_b} and \ref{fig:group_c}, respectively.

In Fig.~\ref{fig:emuvvz0h_yz_yh_30TeV_if_yzcut0}, 
we demonstrated that an appropriate kinematical cut can remove specific contributions
so that partial FD-gauge subamplitudes can describe a physical distribution.  
Although we only discussed the rapidity distributions and imposed the rapidity cut on $y(Z)$ in Eq.~\eqref{yzcut}
as a demonstration in 
\rev{
the main text,
}
other kinematical variables such as the transverse momenta,
\rev{
shown in Appendix,
}
and the invariant masses may further help to extract the VBS contribution, which will be reported elsewhere. 
Such analyses with FD-gauge amplitudes may also help for the validation and understanding of the effective vector-boson approximation or electroweak parton distribution functions for VBF/VBS-type processes,
which have been intensively studied recently; 
see, e.g.  
Refs.~\cite{Borel:2012by,Costantini:2020stv,Han:2020uid,Ruiz:2021tdt,Hamada:2024ojj,Bigaran:2025rvb,Frixione:2025guf,Dahlen:2025udl}. 

In conclusion, we studied the rapidity distributions for the Higgs boson and the Z boson 
for the $l^-l^+\to \nu\bar{\nu}Zh$ process rather in detail with the help of the FD gauge.
Although our studies were limited in the SM, we believe that our findings provide useful information to search for new physics in future experiments. 
In this paper we focused only on the productions of the longitudinally-polarized Z boson.
The basic arguments for the transversely-polarized case are similar, and will be reported elsewhere.
Similar studies can also be applied to VBS $W^\pm h$ productions, where collinear singularity from $t$-channel photon exchange should be carefully taken into account~\cite{Hagiwara:1990gk,Hamada:2024ojj,Hagiwara:2026lul}.

\begin{acknowledgments}
We would like to thank Benjamin Fuks, Kaoru Hagiwara, Junichi Kanzaki, Fabio Maltoni, Olivier Mattelaer, Richard Ruiz, Kei Yamamoto, Ryo Yonamine, and Ya-Juan Zheng for valuable discussions.  
HF also thanks 
the UCLouvain Centre for Cosmology, Particle Physics and Phenomenology
for the hospitality, where parts of this work were performed.
Feynman diagrams are drawn by {\tt TikZ-FeynHand}~\cite{Ellis:2016jkw,Dohse:2018vqo}.
This work is supported in part by JSPS KAKENHI Grants No.~21H01077, No.~21K03583, No.~23K20840, No.~24K07032, and No.~26K00694.
HF is also supported in part by JSPS KAKENHI Grants: the International Leading Research (24K23939).
\end{acknowledgments}

\appendix*
\input pt.tex

\bibliography{bibfd}

\end{document}

%% file: pt.tex
\section{Transverse-momentum distributions}

While we focus on the rapidity ($y$) distributions in the main text,
we show the transverse-momentum ($p_T$) distributions in this appendix as a supplement.

\begin{figure}
  \includegraphics[width=1\columnwidth]{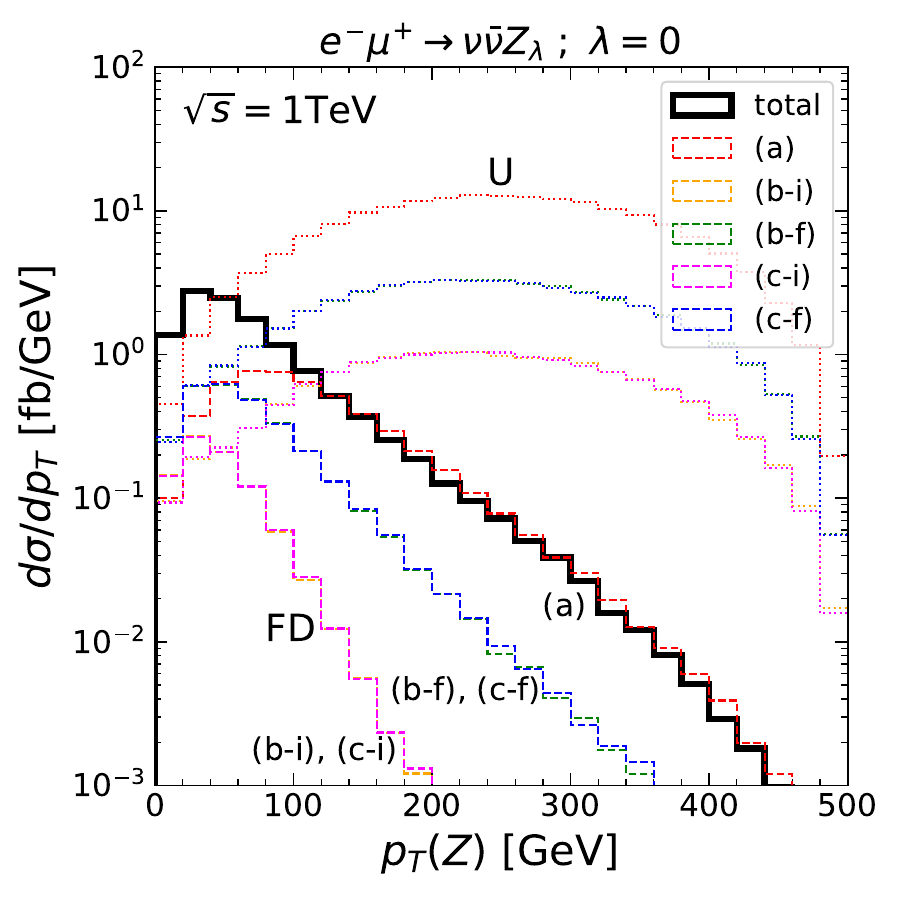}
  \caption{Transverse-momentum distribution of the Z boson for $e^{-}\mu^{+}\to \nu\bar{\nu}Z_{\lambda}$ with $\lambda=0$ at $\sqrt{s}=1$~TeV,
where $\lambda$ is the helicity of the Z boson in the $e^-\mu^+$ collision c.m. frame. 
The solid line denotes the observable distribution, while dashed (dotted) lines show contributions from the absolute value squared of each amplitude, 
as defined in Fig.~\ref{fig:vvz}, in the FD (U) gauge.}
\label{fig:emuvvz0_ptz_1TeV_UvsFD}
\end{figure}

Figure~\ref{fig:emuvvz0_ptz_1TeV_UvsFD} shows the transverse-momentum distribution of the
Z boson for $e^{-}\mu^{+}\to \nu\bar{\nu}Z_{\lambda}$ with $\lambda=0$ at $\sqrt{s}=1$~TeV.
The line styles are the same as those in the rapidity distribution shown in Fig.~\ref{fig:dxsec_FDvsU_l0}.

The $p_T$ distributions of (b-i) and (c-i), which are related
by $y\to -y$, almost completely overlap both in the FD and U gauges, and the same is true for
(b-f) and (c-f).
Thus, the forward-backward separation of the contributions from (b) $e^-W^+$ scattering and (c) $W^-\mu^+$ scattering 
observed in the rapidity distribution in Fig.~\ref{fig:dxsec_FDvsU_l0} is no longer visible in the
$p_T$ distribution.

As in the rapidity distribution in
Fig.~\ref{fig:dxsec_FDvsU_l0}, the shapes from the individual contributions in the U gauge 
differ significantly from the physical distribution.
Especially in the high-$p_T$ region, each U-gauge amplitude gives rise to unphysical energy growth.  
It is therefore difficult to extract the kinematical properties
associated with the individual diagrams from the U-gauge amplitudes.
In contrast, the individual contributions in the FD gauge 
do not exhibit such a large enhancement in the high-$p_T$ region and 
remain comparable in magnitude
to the observable distribution.
The characteristic difference between the FD- and U-gauge amplitudes
discussed for the rapidity distribution is therefore also observed in
the $p_T(Z)$ distribution.

We can then use the individual FD-gauge contributions to examine the
$p_T$ dependence associated with each diagram topology.
The VBF contribution (a) 
nearly overlaps with the observable distribution in the $p_T(Z)>120$~GeV region.
For the contributions of the $Z$-emission from the fermion line, the initial-state emission
contributions (b-i) and (c-i) are more strongly concentrated in the
low-$p_T$ region and decrease rapidly with increasing $p_T$, whereas
the final-state emission contributions (b-f) and (c-f) extend to
higher $p_T$.
This behavior is consistent with the fact that, for initial-state emission, 
the incoming leptons travel along the beam axis, favoring
quasi-collinear emission of the Z boson along the beam direction.
In contrast, for final-state emission, the scattered neutrino can
carry transverse momentum, allowing the emitted Z boson to populate
a broader $p_T$ region.

Such information on the kinematical behavior of the individual
FD-gauge contributions is useful for identifying appropriate
kinematical variables and phase-space regions in which a particular
diagram topology can be enhanced.
In particular, for the
$l^-l^+\to\nu\bar{\nu} Z$ process, since the VBF amplitude (a) contains
the $WWZ$ triple-gauge interaction, this approach can help design
kinematical selections that enhance the VBF contribution and thereby
provide useful information for improving sensitivity to the $WWZ$ coupling.

\begin{figure*}
    \centering
    \includegraphics[width=0.495\textwidth]{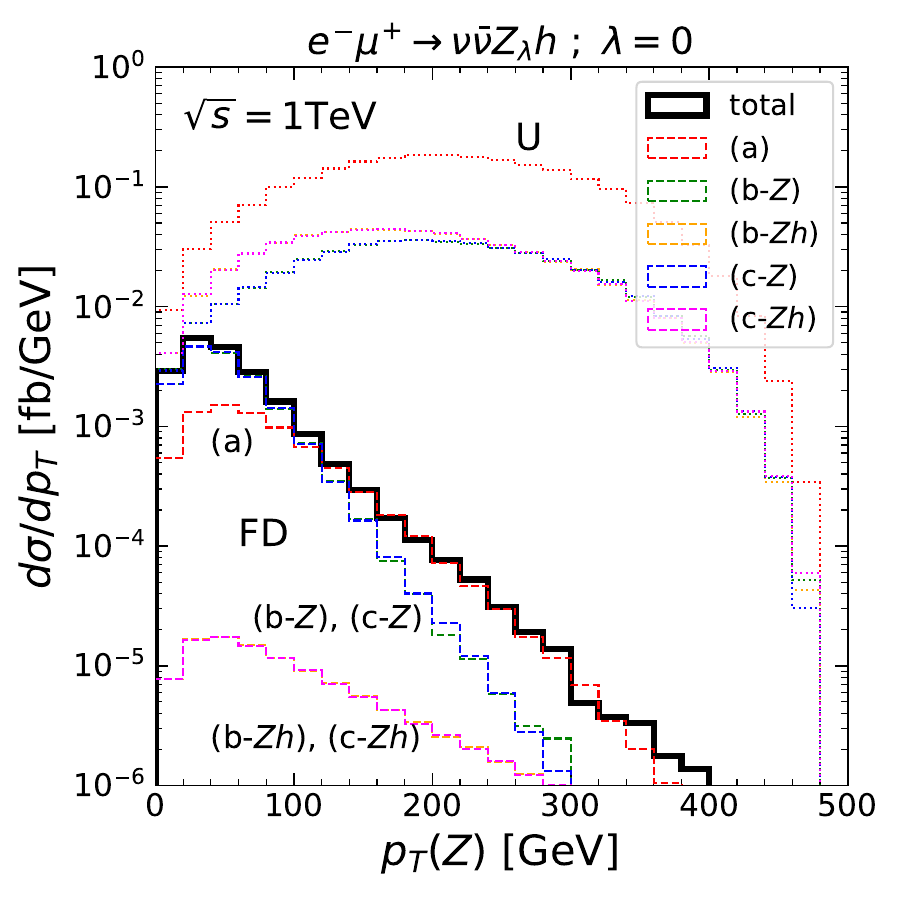}
    \includegraphics[width=0.495\textwidth]{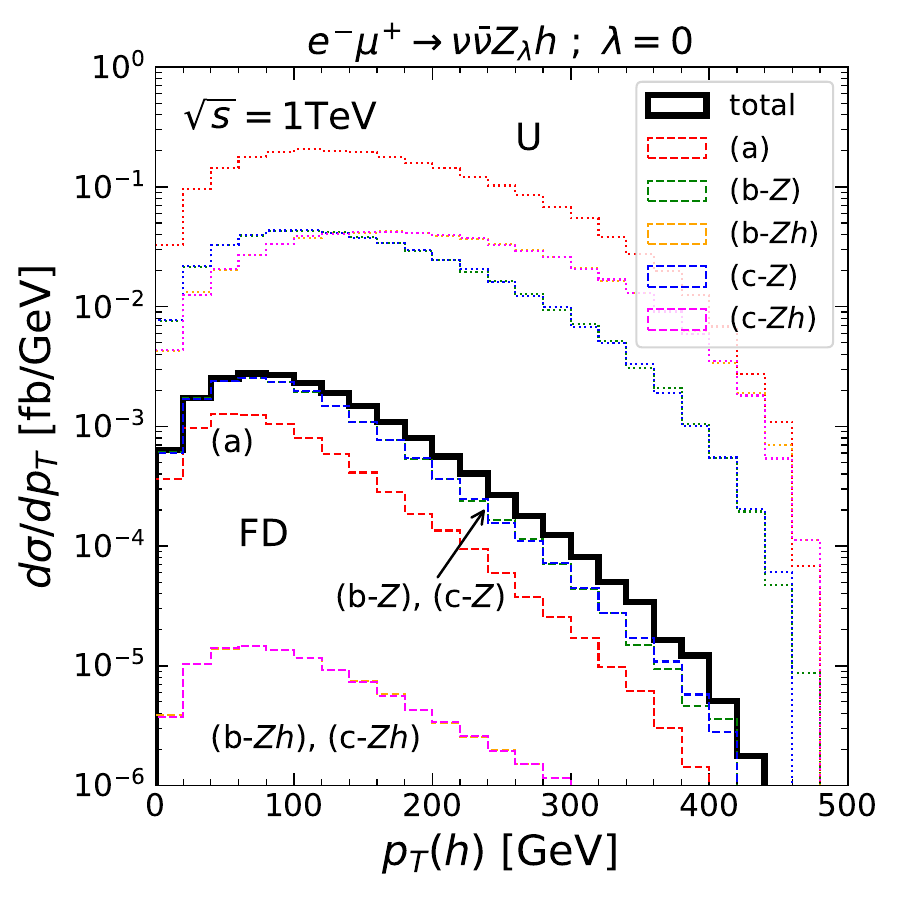}
    \caption{Transverse-momentum distributions of the Z boson (left panel) and the Higgs boson (right panel) for $e^{-}\mu^{+}\to \nu\bar{\nu}Z_{\lambda}h$ with $\lambda=0$ at $\sqrt{s}=1$~TeV,
where $\lambda$ is the helicity of the Z boson in the $e^-\mu^+$ collision c.m. frame. 
The solid line denotes the observable distribution, while dashed (dotted) lines show contributions from each diagram category defined in Figs.~\ref{fig:vvzh}--\ref{fig:group_c} in the FD (U) gauge.}
    \label{fig:emuvvz0h_ptz_pth_1TeV_UvsFD}
\end{figure*}

Figure~\ref{fig:emuvvz0h_ptz_pth_1TeV_UvsFD} shows the transverse-momentum
distributions of the Z boson for $\lambda=0$ and the Higgs boson in
$e^-\mu^+ \to \nu_e\bar{\nu}_\mu Z_\lambda h$
at $\sqrt{s}=1~\mathrm{TeV}$.
The line styles are the same as those in the rapidity distributions shown in Fig.~\ref{fig:vvzh_yz_FDvsU_l0}.
Similar to the $\nu\bar\nu Z$ final state, 
the distributions of (b-$Z$) and (c-$Z$) as well as (b-$Zh$) and (c-$Zh$) almost completely overlap
since the transverse momentum does not distinguish the sign of the beam direction.
While the individual contributions in the
U gauge do not provide any useful information,
the individual contributions in the FD gauge  
can provide clear physics interpretation as discussed below.

We first focus on the $p_T(Z)$ distributions of the individual diagram
categories in the FD gauge, shown in the left panel of Fig.~\ref{fig:emuvvz0h_ptz_pth_1TeV_UvsFD}.
%
%
The (b-$Z$) and (c-$Z$) contributions, corresponding to the Z-boson
emission from the fermion line, are strongly concentrated in the
low-$p_T$ region and decrease rapidly with increasing $p_T$.
This behavior is consistent with the tendency of Z-boson emission
along the fermion line to favor kinematics close to the beam direction.
In contrast, the VBS contribution (a), corresponding to
$W^-W^+ \to Zh$, exhibits a relatively harder $p_T$ distribution
than the (b-$Z$) and (c-$Z$) contributions.
%
The (b-$Zh$) and (c-$Zh$) contributions, in which a $Zh$ pair is emitted
from the fermion line, are significantly smaller than the
single-$Z$-emission contributions (b-$Z$) and (c-$Z$).
This hierarchy is consistent with that observed in the total cross
section in Fig.~\ref{fig:vvzh_xsec_FDvsU_l0} and the rapidity distributions in Fig.~\ref{fig:vvzh_yz_FDvsU_l0}, and with the expectation that
the emission of a $Zh$ pair from a fermion line is more strongly
suppressed than single-$Z$ emission.
Moreover, the $p_T(Z)$ spectrum of the $Zh$-emission contributions is much harder than that of the 
single-$Z$-emission contributions.

We finally turn to the $p_T$ distribution of the Higgs boson, shown in the right panel of Fig.~\ref{fig:emuvvz0h_ptz_pth_1TeV_UvsFD}.
Similar to the rapidity distributions shown in Fig.~\ref{fig:emuvvz0h_yz_yh_30TeV_if}, 
the $p_T(h)$ distribution is rather different from the $p_T(Z)$ distribution;
the $p_T(h)$ spectrum is much harder than the $p_T(Z)$ one.
%
The FD-gauge amplitudes reveal
how the different production mechanisms of the Z and Higgs bosons
in each diagram topology are reflected in their transverse kinematics.

In the VBS contribution (a), both the Z and Higgs bosons are produced
through the $W^-W^+ \to Zh$ subprocess and therefore reflect the
kinematics of the same hard-scattering process.

In contrast, in the (b-$Z$) and (c-$Z$) contributions, the Z boson
is emitted from the fermion line, whereas the Higgs boson is produced through the $W^-W^+\to h$ subprocess.
Consequently, the $p_T(Z)$ and $p_T(h)$ distributions exhibit different
kinematical behaviors in these contributions.
In particular, while the (b-$Z$) and (c-$Z$) contributions in the $p_T(Z)$ distribution are strongly
concentrated in the low-$p_T$ region,
these contributions in the $p_T(h)$ distribution show harder spectra similar to the VBS contribution.

On the other hand, in the (b-$Zh$) and (c-$Zh$) contributions, both
the Z and Higgs bosons are produced through $Zh$-pair emission from
the fermion line.
These contributions therefore have a similar $p_T$ spectrum both in the $p_T(Z)$ and $p_T(h)$ distributions.

Thus, 
in the FD gauge,
we can examine how the different diagram topologies---VBS,
single-$Z$ emission, and $Zh$-pair emission---are reflected in the
transverse kinematics of the Z and Higgs bosons.